\documentclass[%
reprint,
twocolumn,
superscriptaddress,
nofootinbib,
 amsmath,amssymb,
 aps,
]{revtex4-1}
\usepackage{graphicx}
\usepackage{dcolumn}
\usepackage{array}
\usepackage[colorlinks,linkcolor=blue,anchorcolor=blue,citecolor=blue,urlcolor=blue,]{hyperref}
\usepackage{orcidlink}
\usepackage{bm}
\usepackage{algorithm2e}
\usepackage{xcolor}
\usepackage{enumitem}
\usepackage{graphicx}
\usepackage[export]{adjustbox}
\usepackage{units}

\graphicspath{{figures/}}
\begin{document}

\newcommand{\xxf}[1]{\textcolor{purple}{{\it{[XXK: #1]}}}}
\newcommand{\revd}[1]{\textcolor{black}{{{#1}}}}
\newcommand{\changed}[1]{\textcolor{black}{{{#1}}}}
\mathchardef\mhyphen="2D

\preprint{APS/123-QED}

\title{Correcting for Evidence Uncertainty in the Templated Background Search for the Stochastic Gravitational-Wave Background}

\author{Xiao-Xiao Kou\,\orcidlink{0000-0002-7300-370X}}
\email{kou00016@umn.edu}
\affiliation{%
University of Minnesota, School of Physics and Astronomy, Minneapolis, MN 55455, USA}%

\author{Argyro Sasli\,\orcidlink{0000-0001-7357-0889}}
\affiliation{%
University of Minnesota, School of Physics and Astronomy, Minneapolis, MN 55455, USA}%

\author{Muhammed Saleem\,\orcidlink{0000-0002-3836-7751}}
\affiliation{Center for Gravitational Physics, University of Texas at Austin, Austin, TX 78712, USA}%

\author{Vuk Mandic\,\orcidlink{0000-0001-6333-8621}}
\affiliation{%
University of Minnesota, School of Physics and Astronomy, Minneapolis, MN 55455, USA}%
\date{\today}

\begin{abstract}
We report a new source of systematic bias in the Bayesian Templated Background Search (TBS) for the astrophysical stochastic gravitational-wave background. This bias arises from the statistical uncertainty in nested sampling evidence estimates at the segment level and persists even when the likelihood model accurately characterizes the data. Using an analytically tractable model and a frequency-domain mock data analysis with an astrophysically motivated binary black hole population, we show that this uncertainty coherently inflates the inferred duty cycle by up to an order of magnitude when millions of segments are accumulated. We propose a simple and efficient correction and demonstrate its effectiveness across all tested configurations. We also find that a search prior restricted to high optimal signal-to-noise ratio (SNR) misattributes intermediate-strength signals, while extending the prior to support weaker signals amplifies the evidence uncertainty. Both the correction and careful search prior design are essential for robust inference as TBS analyses scale to upcoming observing runs.
\end{abstract}

\maketitle

\section{Introduction}
The LIGO–Virgo–KAGRA (LVK) collaborations~\cite{LIGOScientific:2014pky,VIRGO:2014yos,KAGRA:2018plz} have assembled a catalog of several hundred compact binary coalescences through the end of the O4 observing run~\cite{LIGOScientific:2018mvr,LIGOScientific:2020ibl,LIGOScientific:2021usb,KAGRA:2021vkt,LIGOScientific:2026tep,LIGOScientific:2026wfs} and the catalog is expected to expand substantially with the upcoming O5 observation run due to improved sensitivity~\cite{KAGRA:2013rdx}. These observations enable a growing understanding of the underlying binary black hole (BBH) population~\cite{LIGOScientific:2020kqk,KAGRA:2021duu,LIGOScientific:2025pvj,LIGOScientific:2026ctl} including its mass spectrum, spin distribution, and merger rate evolution with redshift, as well as their formation channels (see, e.g., Refs.~\cite{Mandel:2018hfr,Mapelli:2020vfa,Mapelli:2021taw,Callister:2024cdx,Biscoveanu:2026ikx}). However, these individually resolved events represent less than $1\%$~\cite{KAGRA:2013rdx,KAGRA:2021duu,LIGOScientific:2025pvj,LIGOScientific:2026ctl} of total mergers in the Universe and the remaining ones are too distant and too faint to be directly detected. While invisible on an event-by-event basis, these unresolved signals are not entirely lost: they superpose to form an astrophysical stochastic gravitational-wave background (SGWB) that encodes independent and complementary information about the full BBH population~\cite{Bavera:2021wmw,KAGRA:2021kbb,LIGOScientific:2025kry, Ebersold:2025izh, LIGOScientific:2025bgj}, offering a pathway to probe the merger rate and source properties~\cite{Zhu:2011bd, Mandic:2012pj, Suvodeep:2019, Turbang:2023tjk,Sah:2023bgr,Kou:2024gvp,Lalleman:2024zjt,Sah:2025agw} well beyond the horizon of individual detection.

For the astrophysical background generated by stellar-mass BBH systems, the average duration of GW signals is expected to be shorter than the average interval between consecutive BBH mergers within the frequency band of ground-based detectors~\cite{Johnson:2024foj}. This results in an intermittent or \textit{non-Gaussian} distribution of signals in the data, violating the assumption of a stationary and Gaussian background in typical cross-correlation SGWB searches with LVK data~\cite{Allen:1997ad, Romano:2016dpx}. Therefore, the standard cross-correlation search is suboptimal for the BBH background, leading to reduced search sensitivity and longer time-to-detection. To address this issue, Smith and Thrane~\cite{Smith:2017vfk} proposed leveraging the inherent structure of the signals through the Bayesian Templated Background Search (TBS) framework. This approach offers a significantly reduced time-to-detection compared to traditional cross-correlation methods. The TBS approach divides the continuous data stream into short segments with the same duration, assesses the Bayesian evidence for signal versus noise for each segment, and integrates the resulting evidence ratios within a mixture model by introducing the duty cycle, the fraction of segments containing a BBH signal. This search framework also allows for the simultaneous characterization of the population properties of the BBH sources contributing to the background without imposing a clear separation between the foreground and background~\cite{Smith:2020lkj, Banagiri:2020kqd, Biscoveanu:2020gds, Bers:2025tei}.

Implementing this framework on a large scale, however, reveals subtle yet critical vulnerabilities. 
Progress has been made in modeling and mitigating several key sources of systematic errors, as demonstrated in recent studies~\cite{Talbot:2020auc,Talbot:2021igi,Talbot:2025vth,Kou:2025bhk}.
However, one important aspect that has been largely overlooked has to do with aggregating thousands to millions of segments---even minor inaccuracies in each segment can accumulate, resulting in systematic biases in such large averages. More specifically, the true evidence (or the Bayes factor) is fundamentally inaccessible when stochastic sampling algorithms (e.g., the nested sampling~\cite{Skilling:2004pqw, Skilling:2006gxv,Ashton:2022grj}) are employed to approximate the evidence integral, even when the likelihood model has been optimized to accurately characterize the data. This limitation introduces an irreducible statistical uncertainty in the estimation of the evidence and it affects all segments.

In this work, we investigate how this segment-level evidence uncertainty propagates into a systematic bias on the inferred duty cycle within the TBS framework. We demonstrate the bias through two complementary examples: a simple and closed-form model adapted from~\cite{Essick:2022ojx}, and a mock data challenge utilizing a reduced-dimensional BBH parameter space in which ground-truth evidences are obtainable through optimized grid-based numerical integration. We derive a simple, statistically motivated correction to the mixture-model likelihood and show that it effectively eliminates the bias at negligible computational cost. Beyond the duty cycle, we show that uncorrected evidence uncertainty also distorts the joint inference of BBH population hyperparameters. Finally, we examine how the choice of search prior interacts with the evidence uncertainty problem, revealing a tension between sensitivity to intermediate-strength signals and the growth of per-segment evidence uncertainty at low optimal signal-to-noise ratio (SNR). Our findings establish that accounting for evidence uncertainty and carefully designing the search prior are both essential requirements, not merely a technical refinement for robust duty cycle estimation and population inference within the TBS framework.

This paper is organized as follows. In Sec.~\ref{sec:bias_origin}, we briefly review the mixture-model framework and identify how statistical uncertainty in the segment-level evidence enters the likelihood. In Sec.~\ref{sec:toy_model}, we demonstrate the resulting bias on the duty cycle through a toy model and present a simple correction to the mixture-model likelihood to fix the bias. In Sec.~\ref{sec:tbs_mock}, we validate this correction using a mock data challenge based on a reduced-dimensional BBH parameter space with an astrophysically motivated population model, and examine the implications for searches targeting weaker signals and the design of the search prior. We summarize our findings and conclude in Sec.~\ref{sec:summary}, and offer further details on our procedures in the Appendices. 

\section{Intermittent Search and The origin of Bias}
\label{sec:bias_origin}
\subsection{The Mixture Model}
The unresolved astrophysical background generated by stellar-mass BBHs is characterized by its intermittent nature in the frequency band of ground-based detectors, as it consists of ``popcorn-like" signals separated by extended periods of pure noise~\cite{LIGOScientific:2017zlf}. Intermittency is typically quantified by a metric known as the duty cycle $\xi$, which represents the fraction of data segments containing a GW signal. To effectively search for these discrete, intermittent signals, we can divide the continuous detector data strain into short, non-overlapping segments~\cite{Drasco}, typically lasting 4 seconds so as to ensure that the probability of observing multiple overlapping signals in a single segment is negligibly small. The data in each segment $d_i$ is then evaluated using a hierarchical mixture likelihood
\begin{equation}
\label{eq:simple_mixture}
\mathcal{L}\left(d_i\,|\, \xi\right)=\xi \mathcal{Z}_s^i + (1-\xi) \mathcal{Z}_n^i,
\end{equation}
where $\mathcal{Z}_n^i=\mathcal{L}(d_i\,|\,N)$ is the noise evidence and $\mathcal{Z}_s^i = \mathcal{L}(d_i\,|\,S)$ is the signal evidence, obtained by marginalizing the likelihood over the signal parameters $\theta$ with respect to a prior $\pi(\theta)$,
\begin{equation}
\label{eq:signal_evidence}
\mathcal{Z}_s^i = \int \mathcal{L}(d_i\,|\,h(\theta))\,\pi(\theta)\,d\theta,
\end{equation}
where $h(\theta)$ denotes the signal template evaluated at parameters $\theta$. The choice of template determines how the signal is modeled in each segment and is the key distinction between different implementations of the mixture model framework.

Information across the entire observation period is then stacked by taking the product of the individual mixture likelihoods from Eq.~\ref{eq:simple_mixture} over $N$ independent segments $\mathcal{D}=\{d_i\}$
\begin{equation}
\label{eq:total_mixture}
\mathcal{L}\left(\mathcal{D}\,|\, \xi \right) = \prod_i^N \left( \xi \mathcal{Z}_s^i + (1-\xi) \mathcal{Z}_n^i \right).
\end{equation}
Two principal approaches have been developed within this framework, differing in their choice of signal template. The deterministic TBS~\cite{Smith:2017vfk} uses compact binary chirp waveforms as the template $h(\theta)$, evaluating the likelihood directly at the strain level with $\theta$ representing the physical parameters of the binary (masses, spins, luminosity distance, sky locations, etc.). The stochastic search for intermittent GWBs (SSI)~\cite{Lawrence:2023buo} instead models the signal as stochastic bursts, using a power spectrum template to evaluate excess cross-correlated power between detectors. Despite these methodological differences, both approaches fundamentally rely on the same mixture model structure of Eqs.~\ref{eq:simple_mixture}--\ref{eq:total_mixture} and therefore both require an accurate evaluation of the evidence integral in Eq.~\ref{eq:signal_evidence} for each individual data segment. While our analysis and mock data studies are conducted within the TBS framework, the bias mechanism and the correction we propose may apply equally to SSI or any other search built on this mixture model, as the source of the bias lies in the evidence estimation rather than in the specific form of the template.

\subsection{Locating Biases and Uncertainties}
\label{subsec:source_bias}
Both the TBS and SSI frameworks must accumulate millions of independent data segments to extract information from weak signals, but this massive accumulation presents a severe challenge: small inaccuracies in the likelihood model at the individual segment level are compounded when multiplied across segments, leading to significantly biased results. Significant progress has been made in identifying and mitigating several of these data-driven systematic biases. For TBS, these include the effects of non-Gaussian noise transients, uncertainties in the noise power spectral density, and correlation between frequency bins from the finite duration of each segment~\cite{Smith:2017vfk,Talbot:2020auc,Talbot:2021igi,Kou:2025bhk}. For SSI, the progress has focused on a more careful treatment of the overlap reduction function~\cite{Liu:2026xhc}.

However, aside from the likelihood modeling, it is also important to note that the segment-level signal evidences (Eq.~\ref{eq:signal_evidence}) involve analytically intractable integrals over a multidimensional parameter space. For computational considerations, both TBS and SSI rely heavily on stochastic numerical algorithms, such as nested sampling (NS), to approximate them. The estimates of $\mathcal{Z}^i_s$ by a nested sampling algorithm are subject to statistical and systematic errors. Consequently, the analysis output from one specific data segment should be regarded as a random variable rather than a deterministic quantity. This effect has been assumed to be small and was typically ignored in previous analyses~\cite{Smith:2017vfk,Smith:2020lkj,Banagiri:2020kqd,Lawrence:2023buo,Kou:2025bhk,Liu:2026xhc}. We note, also, the remark by~\cite{HernandezVivanco:2019fku} that the accuracy of estimating the evidence is closely related and critical to the choice of the prior. More recently, Ref.~\cite{Bers:2025tei} raised the possibility that biases observed when restricting the analysis to the weakest signals could originate in evidence estimation error, though the effect was neither quantified nor corrected.

The estimated log-evidence $\ln \tilde{\mathcal{Z}}_i$ in nested sampling is typically modeled~\cite{Skilling:2006gxv,Speagle:2019ivv} as a Gaussian random variable centered on the unknown true value $\ln \mathcal{Z}_i$ with its standard deviation $\sigma_i$ 
\begin{equation}
\label{eq:ns_uncertainty}
\ln \tilde{\mathcal{Z}}_i \sim \mathcal{N}(\ln \mathcal{Z}_i,\;\sigma^2_i), \quad \sigma_i \sim \sqrt{\frac{H_i}{n_{\rm{live}}}},
\end{equation}
where $n_{\rm{live}}$, referred to as the number of live points, is an important configuration parameter to initialize a nested sampling run, while $H_i$ represents the information gain (or Kullback-Leibler divergence~\cite{Kullback:1951zyt}) between the prior and posterior. This relationship arises because the dominant source of statistical uncertainty is the Poisson variability inherent in the stochastic shrinkage of the prior volume at each sampling step. Although precisely characterizing this exact error distribution is a highly nontrivial problem, theoretical analyses (see~\cite{Skilling:2006gxv,Higson2017SamplingEI,Higson:2018cqj,Speagle:2019ivv,Fowlie:2022jls} for more details) together with numerical simulations~\cite{Fowlie:2022jls} validate that this formulation is a remarkably robust and accurate approximation.

For real data analysis, the duty cycle is expected to be exceedingly small ($ \lesssim 0.01$), meaning that the overwhelming majority of data segments are pure noise. In this regime, the log Bayes factor $\ln \rm{BF}_i \equiv \ln \mathcal{Z}^i_s - \ln \mathcal{Z}^i_n$ for nearly every segment fluctuates around zero, governed by noise variance as well as the statistical uncertainty inherent in the nested sampling estimate itself. Unlike systematic errors in the likelihood model, which can in principle be improved by more accurate waveform or noise modeling, this stochastic scatter is an intrinsic property of the sampling algorithm itself and persists unless it can be completely eliminated by making $n_{\rm{live}}$ infinite. 
The problem is further exacerbated when the search prior, which encodes our tentative belief about the source population, extends to cosmological distances (e.g., $z \gtrsim 7$). In this regime, the prior supports a large number of weak, deeply sub-threshold signals whose contribution to the data is nearly indistinguishable from pure noise fluctuations on a per-segment basis. The statistical uncertainty in the log Bayes factor estimation compounds this difficulty, as it further blurs the already marginal distinction between noise-only segments and those containing weak signals, making it even harder for the mixture model to correctly infer the true duty cycle.

\section{Demonstrating and Correcting the Bias}
\label{sec:toy_model}
\subsection{Univariate-Gaussian Toy Model}
\label{subsec:toy_model}
To isolate and illustrate the impact of evidence uncertainty on duty cycle inference, we construct a minimal toy model in which all relevant quantities, including the signal and noise evidences, the Bayes factor and the information gain, admit closed-form expressions. 
While we do not expect this toy model to quantitatively reproduce the full $\ln \rm{BF}_i$ distribution of a realistic TBS analysis, it does capture the essential feature that the noise and signal distributions strongly overlap and have extended tails, which is the regime where evidence uncertainty has the greatest impact. A quantitative validation using a realistic BBH signal model is presented in Sec.~\ref{sec:tbs_mock}.

\begin{figure}[t]
    \centering
    \includegraphics[width=1.0\linewidth]{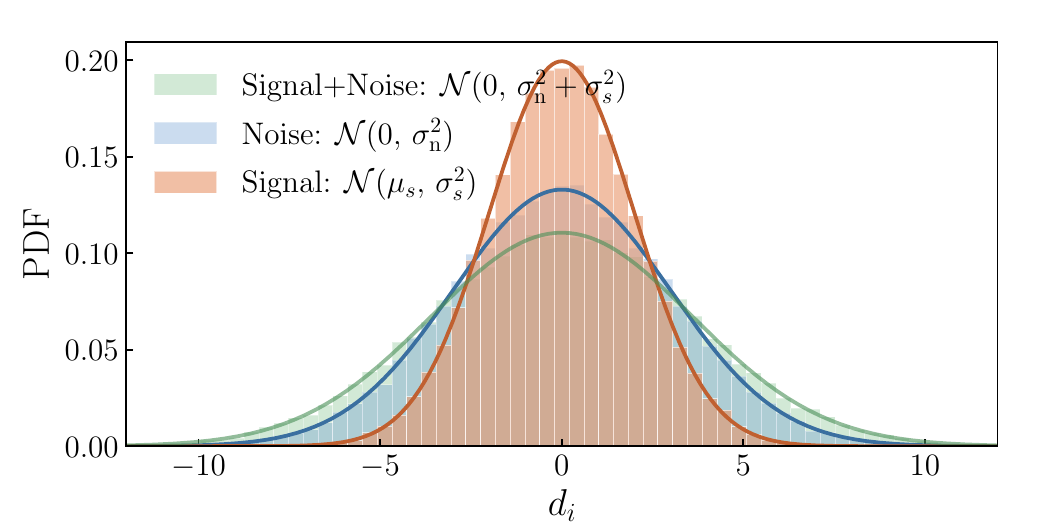}
    \caption{Distributions of $d_i$ in the univariate-Gaussian model (Eq.~\ref{eq:toy_model}) with $\mu_s=0, \sigma_s=2$ and $\sigma_{\rm{n}}=3$. The orange curve shows the signal-prior distribution, the blue curve shows the noise distribution and the green curve shows the marginalized signal-plus-noise distribution.}
    \label{fig:toy_distributions}
\end{figure}

Following the approach of~\cite{Essick:2022ojx}, we consider a univariate setting in which each data segment $d_i$ consists of a single scalar draw, with both signals and noise following Gaussian distributions:
\begin{equation}
\label{eq:toy_model}
x_s \sim \mathcal{N}(\mu_{s}, \sigma^2_{s}), \;x_{\mathrm{obs}} \sim \mathcal{N}(x_s,\;\sigma^2_{\rm{n}}),
\end{equation}
where we have chosen $\mu_s=0, \sigma_s=2$ and $\sigma_{\rm{n}}=3$ such that the typical SNR is modest, mimicking the regime in which the majority of astrophysical BBH signals are deeply sub-threshold. As derived in Appendix~\ref{app:toy_details}, these parameters yield expected log Bayes factors of $\mathbb{E}[\ln \rm{BF}] \approx -0.03$ for noise segments and $\approx 0.04$ for segments having signals, with both distributions concentrated near zero and substantially overlapping.  
The analytic expressions for the signal and noise evidences, the resulting log Bayes factor, and the information gain are straightforward to evaluate in this setting; we collect them in Appendix~\ref{app:toy_details} for completeness.

In Fig.~\ref{fig:toy_distributions}, we present the resulting distributions of $d_i$ under three hypotheses: noise-only, signal-only, and signal-plus-noise. The significant overlap between the noise and signal-plus-noise distributions highlights the fundamental difficulty of this problem: most segments that contain a signal yield data statistically indistinguishable from pure noise on a per-segment basis, and information about the duty cycle must therefore be extracted from the collective statistical properties of the full ensemble.

\begin{figure}[t]
    \centering
    \includegraphics[width=1.0\linewidth]{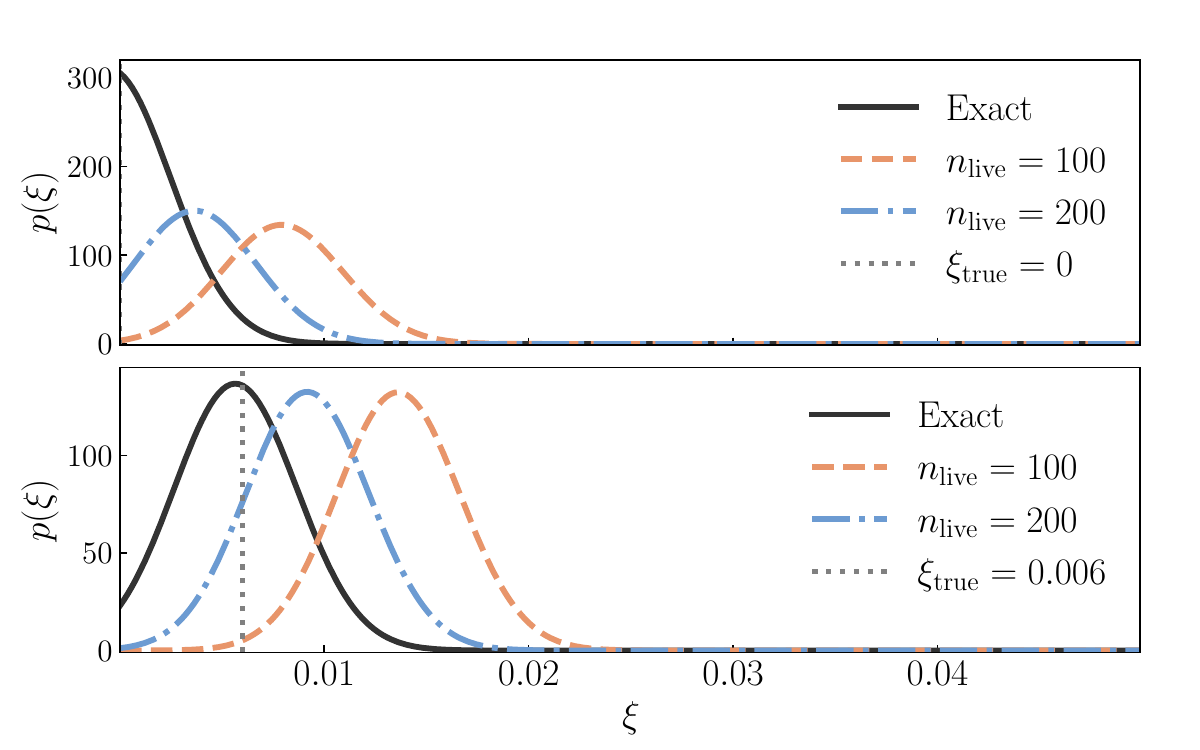}
    \caption{Posterior distribution of the duty cycle $\xi$ inferred from the univariate-Gaussian model. The top panel corresponds to a pure-noise dataset ($\xi_{\rm{true}}=0$) and the bottom panel to a dataset with a simulated duty cycle of $\xi_{\rm{true}}=0.006$. In each panel, the solid black curve shows the posterior obtained using the exact, analytically computed log Bayes factor, while the dashed orange and dash-dotted blue curves show the posterior when the log Bayes factors are perturbed by Gaussian noise with standard deviation $\sigma_i = \sqrt{H_i/n_{\rm{live}}}$ for $n_{\rm{live}}=100$ and $n_{\rm{live}}=200$, respectively.}
    \label{fig:toy_bias_uncorrected}
\end{figure}

Utilizing this toy model, we generate mock datasets for two representative duty cycles $\xi=0$ (pure noise) and $\xi=0.006$, comprising one million data segments, and analyze each using the mixture model of Eq.~\ref{eq:total_mixture}. To emulate the effect of evidence estimation uncertainty due to a finite number of live points used in nested sampling, we perturb the analytically computed log Bayes factor of each segment with a Gaussian random variate. The standard deviation for this perturbation is taken from Eq.~\ref{eq:ns_uncertainty} together with $H_i$ derived in Eq.~\ref{eq:toy_KL}, and we conduct the analysis for two different numbers of live points\footnote{Note that $n_{\rm{live}}=100$ and $200$ represent our optimistic choices for a one-dimensional problem. In standard BBH parameter estimation with a full 15-dimensional parameter space, typical analyses employ $n_{\rm{live}}\sim 1000-2500$. Since the evidence uncertainty scales as $\sigma \sim \sqrt{H/n_{\rm{live}}}$ and the information gain $H$ grows with the dimensionality of the parameter space, the bias demonstrated here is expected to be further amplified in realistic high-dimensional settings.}: $n_{\rm{live}}=100$ and $200$. The results, shown in Fig.~\ref{fig:toy_bias_uncorrected}, reveal a clear and systematic upward bias in the recovered duty cycle when evidence uncertainty has not been properly modeled. For $\xi=0$, when no signals are present, ignoring evidence estimation uncertainty leads to a nonzero duty cycle recovery. This bias is more pronounced for smaller $n_{\rm{live}}$, consistent with the larger per-segment scatter, and persists at $\xi=0.006$, where it distorts the recovery of the true simulated value. These results confirm that the evidence uncertainty does not average out over a large number of segments but instead coherently shifts the posterior on $\xi$.

\subsection{Introducing the Correction}
The bias demonstrated above arises because the standard mixture model in Eq.~\ref{eq:total_mixture} treats the estimated evidence from each segment as though it were exact, neglecting the systematic upward shift introduced by the log-normal nature of the evidence uncertainty. Since the log-evidence error is approximately Gaussian, $\epsilon_i \sim \mathcal{N}(0, \sigma^2_i)$, the estimated Bayes factor $\tilde{\rm{BF}}_i$ relative to the true one $\rm{BF}^*_i$ can be written as $\tilde{\rm{BF}}_i=\rm{BF}^*_i\exp{(\epsilon_i)}$. Its expectation value $\mathbb{E}[\tilde{\rm{BF}}_i]=\rm{BF}^*_i\exp{(\sigma_i^2/2)}$ is then biased high relative to the true value by a factor of $\exp{(\sigma_i^2/2)}$.

A natural correction\footnote{Another alternative approach would be to marginalize over the unknown true log Bayes factor by treating the NS estimate as a noisy observation. However, this requires specifying the full distribution of the true $\ln \rm{BF}^*_i$ across segments, which depends on the specific signal population prior and noise properties. In contrast, the correction adopted here relies only on the per-segment uncertainty $\sigma_i$, which is a standard output of nested sampling algorithms supported by most packages~\cite{Handley:2015vkr,Williams:2021qyt, Speagle:2019ivv,Karamanis:2022ksp}, and does not introduce any additional model assumptions.} is therefore to subtract $\frac{1}{2}\sigma_i^2$ from each estimated $\ln \tilde{\rm{BF}}_i$ before it enters the mixture likelihood, yielding the unbiased corrected Bayes factor $\hat{\rm{BF}}_i$
\begin{equation}
\label{eq:corrected_bf}
\ln\hat{\mathrm{BF}}_i = \ln\tilde{\mathrm{BF}}_i - \frac{1}{2}\sigma_i^2.
\end{equation}
This correction removes the leading-order bias by ensuring that $\mathbb{E}[\hat{\mathrm{BF}}_i]=\rm{BF}^*_i + \mathcal{O}(\sigma^4_i)$, rather than $\rm{BF}^*_i\exp{(\sigma_i^2/2)}$. The corrected per-segment mixture likelihood then takes the form
\begin{equation}
\label{eq:corrected_mixture}
\mathcal{L}_{\rm corr}(d_i\,|\,\xi) \propto \xi\,\hat{\mathrm{BF}}_i + (1-\xi),
\end{equation}
where $\hat{\mathrm{BF}}_i = \exp{(\ln \tilde{\rm{BF}}_i-\frac{1}{2}\sigma^2_i)}$. As we derive in detail in Appendix~\ref{app:derivation}, the residual bias to the duty cycle after applying this correction is suppressed by an additional factor of true duty cycle $\xi^*$ relative to the uncorrected case. Since $\xi^*$ is expected to be $\lesssim 0.01$ in the TBS context, the correction is highly effective precisely in the regime of greatest practical interest. The correction is also computationally negligible, as it requires only the per-segment uncertainty estimate $\sigma_i$ that is cheap to compute and save. We show in Appendix~\ref{app:validation} that the alternative strategy of simply increasing $n_{\rm live}$ to suppress the bias is both computationally impractical at the scale of millions of segments and insufficient to fully eliminate it.

\begin{figure}[t]
    \centering
    \includegraphics[width=1.0\linewidth]{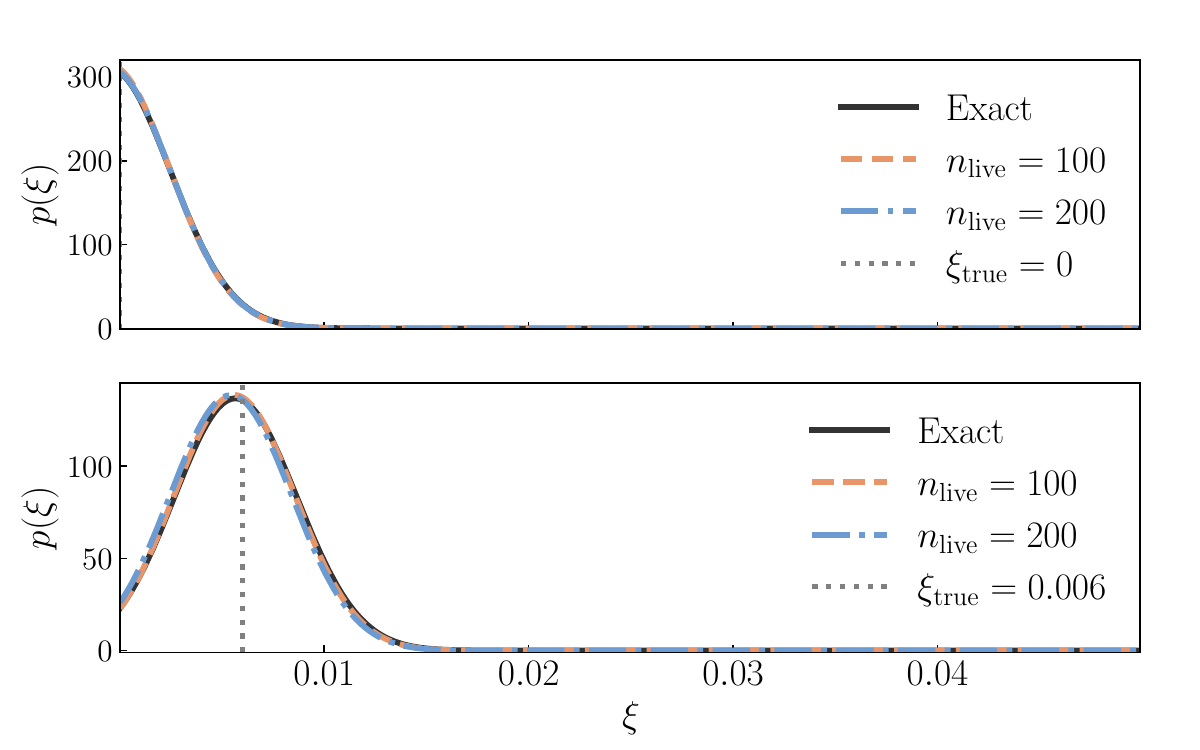}
    \caption{Same as Fig.~\ref{fig:toy_bias_uncorrected}, but introducing the correction for each segment. For both $\xi_{\rm{true}}=0$ (top) and $0.006$ (bottom), the posterior obtained with the same mock data are in excellent agreement with the exact result, confirming that the correction successfully removes the systematic bias without introducing additional free parameters.}
    \label{fig:toy_bias_corrected}
\end{figure}

We apply this corrected likelihood to the same toy-model datasets analyzed in the previous analysis. Fig.~\ref{fig:toy_bias_corrected} shows the resulting posteriors on $\xi$ for the same two simulated values ($\xi_{\rm{true}}=0$ and $0.006$) and the same two choices of live points ($n_{\rm{live}}=100$ and $200$). In contrast to the uncorrected results of Fig.~\ref{fig:toy_bias_uncorrected}, the posteriors obtained with perturbed log Bayes factors after corrections now match the exact result in all four cases. For $\xi_{\rm{true}}=0$, the corrected analysis correctly recovers a posterior that peaks at zero, eliminating the spurious nonzero duty cycle seen in the uncorrected case. For $\xi_{\rm{true}}=0.006$, the corrected posteriors recover the simulated value without the systematic shift toward higher $\xi$ that was evident in Fig.~\ref{fig:toy_bias_uncorrected}. The agreement between the exact and corrected posteriors holds for both values of $n_{\rm{live}}$, confirming that the simple subtraction of $\frac{1}{2}\sigma_i^2$ is sufficient to remove the bias across different levels of sampling noise.

The analysis above has focused on the case where the duty cycle $\xi$ is the only free parameter, with the population parameters held fixed. In practice, however, the TBS framework is designed to jointly infer $\xi$ together with the population hyperparameters (e.g., $\mu_s$ and $\sigma_s$ in our toy model) that govern the signal distribution. In Appendix~\ref{app:pop_bias}, we extend the toy model analysis to this joint inference setting and show that uncorrected evidence uncertainty not only biases the duty cycle but also distorts the recovered population hyperparameters through their degeneracy with $\xi$. The correction of Eq.~\ref{eq:corrected_bf} restores agreement with the exact posterior across all parameters simultaneously, confirming its effectiveness in the broader context of hierarchical population inference.

\section{TBS Mock Data Analysis}
\label{sec:tbs_mock}
Having demonstrated the bias and its correction in a univariate-Gaussian model, we now proceed to verify that the same effect persists and the correction remains effective in a mock data analysis that captures the essential features of a full TBS analysis. In a real analysis, the target quantity for each segment is the Bayes factor $\rm{BF}_i$, which requires evaluating the evidence integral of Eq.~\ref{eq:signal_evidence} over the 15 parameters characterizing a quasicircular BBH. Unlike the toy model of Sec.~\ref{sec:toy_model}, the gravitational waveform is a highly nonlinear function of these parameters, and the resulting likelihood surface admits no closed-form marginalization. In practice, the evidence must therefore be approximated numerically, and nested sampling has become the dominant algorithm for this task in gravitational-wave data analysis~\cite{Skilling:2004pqw, Skilling:2006gxv, Ashton:2022grj}, with widely used implementations such as \texttt{dynesty}~\cite{Speagle:2019ivv}, \texttt{PyMultinest}~\cite{Buchner:2014nha}, \textsc{nessai}~\cite{Williams:2021qyt}, and \textsc{polychord}~\cite{Handley:2015vkr}, among others. Running such an algorithm on each segment typically requires several hours, and even then the output is only a stochastic estimate of the true evidence. In other words, running nested sampling on millions of segments is not only computationally prohibitive but also precludes access to ground-truth Bayes factors that are critical for this study.

To overcome both limitations, we develop a reduced-dimensional BBH signal model\footnote{We do not follow the approach in~\cite{Renzini:2024hiu} where parameters such as the primary mass are fixed and all binaries are assumed to be in the optimal detector configuration. Fixing the mass eliminates the contribution of the chirp mass integral to the Bayes factor, while assuming optimal orientation removes the signal amplitude modulation due to the detector response. Both simplifications alter the statistical structure of the $\ln \rm{BF}_i$ distribution across segments.} inspired by~\cite{Finn:1992xs} that makes the evidence integral amenable to direct numerical integration while preserving the marginalization over all physically relevant parameters. This approach retains the physical spread in signal strength across the population and allows us to obtain ground-truth log Bayes factors at a fraction of the cost of nested sampling, enabling rapid simulation and analysis of datasets containing millions of segments.

\subsection{Population and Signal Model}
\label{subsec:setup}
We consider a population of non-spinning, equal-mass BBH mergers described by the \textit{Power Law + Peak} (PLPP) mass function~\cite{Talbot:2018cva,KAGRA:2021duu}. The merger rate evolution is modeled by a parametrized Madau-Dickinson form, convolving the star formation rate~\cite{Madau:2016jbv} with formation-to-merger time delay. The time delay distribution is assumed to be log-uniform between the minimum delay of $t_{\rm{min}}=100\,\rm{Myr}$~\cite{Wu:2011ac,Ebersold:2025izh} and extending out to redshift $z=7$. Gravitational waveforms in 4s duration are generated in the frequency domain using the \textsc{IMRPhenomXAS} approximant~\cite{Pratten:2020fqn} with \textsc{WF4Py}~\cite{Iacovelli:2022bbs} for a single LIGO detector while assuming Gaussian noise with O4 sensitivity~\cite{LIGO-T2200043}. For a single detector and a non-precessing, equal-mass and dominant-mode waveform, the antenna pattern functions and inclination angle are absorbed into a single effective amplitude factor $\Theta$. We describe this reduction factor and the resulting likelihood, the Bayes factor, and the KL divergence expressions in detail in Appendix~\ref{app:tbs_likelihood}.

Recent work~\cite{Renzini:2024hiu} has shown that templated background searches for the BBH background gain most of their information from the low-redshift region of the population, in contrast to conventional cross-correlation methods. However, redshift alone is not a sufficient proxy for signal strength in this context, as the intrinsic loudness of a signal is collectively determined by both the redshifted chirp mass and the luminosity distance, which together set the optimal SNR $\rho_{\rm{opt}}$ (assuming an optimally oriented and located source). The effective detector response $\Theta$ modulates all signals through the same distribution regardless of their intrinsic parameters. We note that another work~\cite{Bers:2025tei} recently studied the impact of detector matched-filter SNR from selected data segments on duty cycle inference and showed that dominant information obtained by the TBS on the duty cycle inference comes from segments with matched-filter SNR $\sim 5\text{--}7$. While matched-filter SNR is the relevant quantity from a data-driven perspective, it folds in the detector response to each individual source, so that intrinsically loud signals can appear weak due to an unfavorable sky location or inclination. This makes it difficult to disentangle the intrinsic contribution of different parts of the population to the duty cycle, as signals from the same optimal SNR bin can scatter across a wide range of detected SNR values. Optimal SNR, by contrast, isolates the intrinsic signal strength and therefore provides a more natural basis for organizing the source population, offering a direct probe into both the information content of each signal and its associated evidence uncertainty.

\begin{figure}[t]
    \centering
    \includegraphics[width=1.0\linewidth]{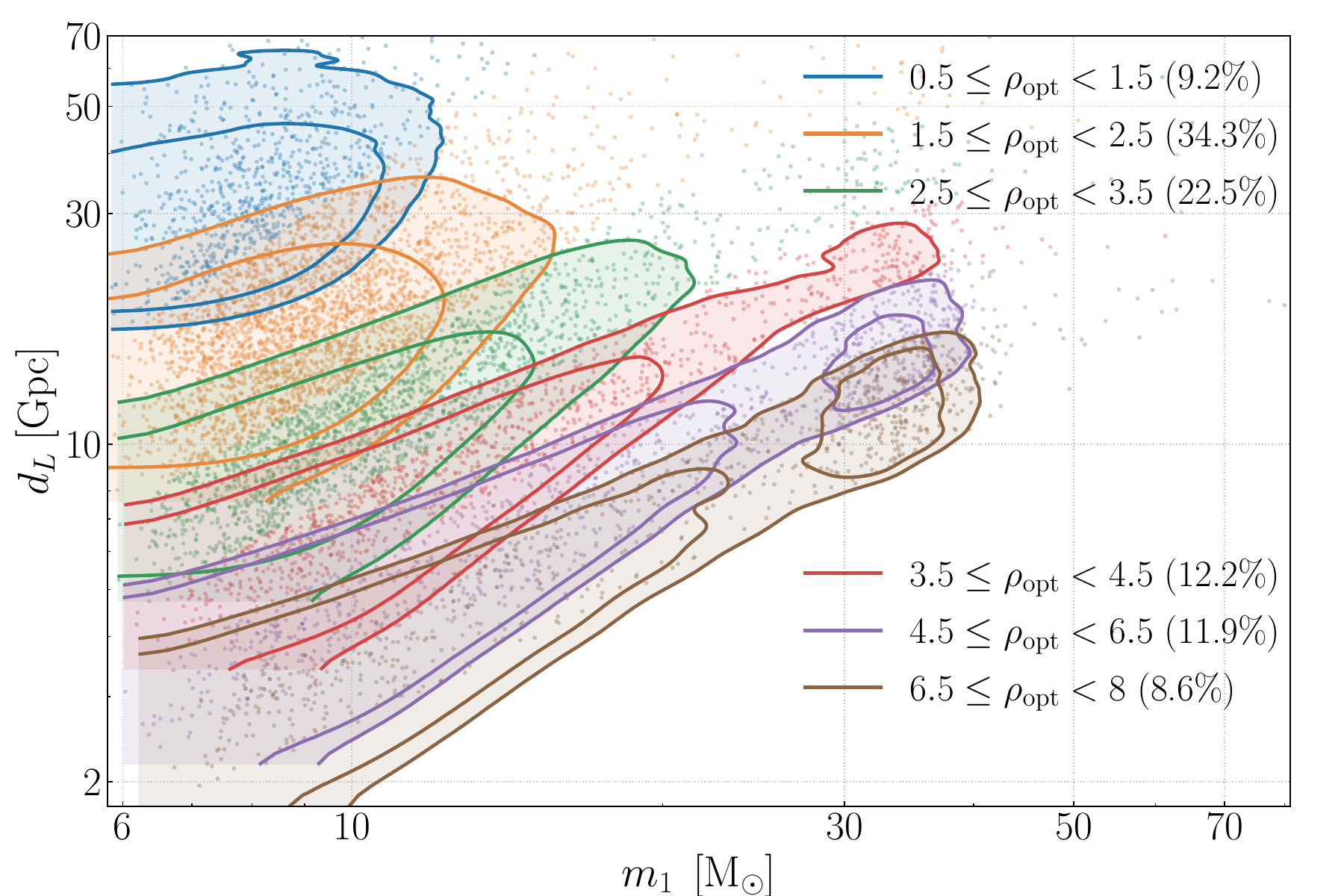}
    \caption{Partition of the BBH source population in parameter space of the source-frame primary mass $m_1$ and luminosity distance $d_L$, color-coded by optimal SNR bin. Contours enclose the $68\%$ and $90\%$ credible regions for each bin, with scatter points showing representative samples. The percentages in the legend indicate the fraction of the total population contained in each bin. The majority of sources ($\sim 34\%$) fall in the $1.5 \leq \rho_{\rm{opt}} < 2.5$, with progressively smaller fractions at higher optimal SNR. Sources in the lowest SNR bins occupy the largest volume in this parameter space, extending to the highest redshifts and luminosity distances. The population is drawn from the PLPP mass function combined with a Madau-Dickinson merger rate with time delay, assuming non-spinning, equal-mass-ratio binaries and O4 detector sensitivity.}
    \label{fig:snr_partition}
\end{figure}

To this end, we partition the intrinsic $(m_1, d_L)$ space into bins of optimal SNR spanning ranges such as $0.5 \leq \rho_{\rm{opt}} < 1.5$ up to $6.5 \leq \rho_{\rm{opt}} < 8$, as illustrated in Fig.~\ref{fig:snr_partition}. The figure shows that the majority of the population ($\sim 34\%$) falls in the $1.5 \leq \rho_{\rm{opt}} < 2.5$ bin, with progressively smaller fractions at higher optimal SNR, confirming that deeply sub-threshold signals dominate the astrophysical population. The purpose of this partition is twofold: it allows us to study how the evidence uncertainty and its impact on the duty cycle depend on the intrinsic signal strength, and it provides the basis for a further simplification. Fig.~\ref{fig:snr_partition} is shown in the source-frame variables $(m_1, d_L)$ in which the mass function and merger rate are defined, while the signal amplitude depends on the detector-frame chirp mass $\mathcal{M}_c^{\rm{det}}$ and the luminosity distance, which together set $\rho_{\rm{opt}}$. Each bin therefore maps to a distribution of detector-frame chirp mass, from which we draw $\mathcal{M}_c^{\rm{det}}$ for each source. We then assign to each bin a representative optimal SNR
$\rho_{\rm{opt}} = 1.3, 2, 3, 4, 5.3$ and $6.5$, taken from the medians, and adjust the luminosity distance that fixes the optimal SNR of every source at $\rho_{\rm{opt}}$, which reduces the evidence integral to a single quadrature over $\mathcal{M}_c^{\rm{det}}$ (see Appendix~\ref{app:tbs_likelihood}). This construction redistributes the luminosity distances slightly relative to the original population, since $\rho_{\rm{opt}}$ does not exactly match the optimal SNR of every source in the bin. The effect is small because each bin spans a narrow range of optimal SNR, and we have checked that it does not appreciably alter the overall population structure. Each bin characterized by $\rho_{\rm opt}$ defines a search prior as well as a subset of the population: analyzing a bin means marginalizing the evidence integral over the chirp mass distribution and detector response it supports. A search configuration may use a single bin or the union of several.


Using this reduced-dimensional model, we construct mock datasets as follows. We generate $1.25$ million 4-second pure-noise segments in the frequency domain and compute the exact $\ln \rm{BF}_i$ and KL divergence $H_i$ for each segment by direct grid-based numerical integration under the search prior of each optimal SNR bin. For signal segments, we generate 20,000 injections per SNR bin by drawing the detector-frame chirp mass from the population model, assigning the luminosity distance to match the representative $\rho_{\rm{opt}}$ of that bin, and drawing the effective detector response $\Theta$ from its prior (see Appendix~\ref{app:tbs_likelihood} for modeling details). The exact $\ln \rm{BF}_i$ and $H_i$ are computed for each signal segment in the same manner. Mock datasets are then assembled by specifying a simulated total duty cycle $\xi_{\rm{tot}}$ and a total number of segments $N_{\rm{tot}}$, with the fraction of signals in each SNR bin held fixed at the ratios determined by the population model, so that varying $\xi_{\rm{tot}}$ scales all bins uniformly while preserving the relative population composition.

\subsection{Modeling Nested Sampling Uncertainty}
A key ingredient of our mock data analysis is the incorporation of realistic uncertainty in the per-segment log Bayes factors. As established in Sec.~\ref{subsec:source_bias}, this uncertainty is characterized by $\sigma_i \approx \sqrt{H_i/n_{\rm{live}}}$ (Eq.~\ref{eq:ns_uncertainty}). While this relation provides a direct way to assign $\sigma_i$ given $n_{\rm{live}}$, the mapping between the two depends sensitively on the structure and dimensionality of the parameter space. In particular, the information gain $H_i$ depends on both the data and the choice of prior, so the same $n_{\rm{live}}$ can yield significantly different uncertainties in different analyses. Since our reduced model operates in a three-dimensional parameter space $(m_1, d_L, \Theta)$ rather than a full 15-dimensional TBS analysis, directly adopting a fixed value such as $n_{\rm{live}}=1000$ would not faithfully reproduce the uncertainty encountered in practice. 

To calibrate a realistic level of uncertainty, we use results from a previous TBS analysis~\cite{Kou:2025bhk}, in which full parameter estimation was performed on several thousand segments with $n_{\rm{live}}=1000$, providing empirical measurements of $\sigma_i$ under realistic conditions. Since the prior used in that analysis differs from the one adopted here, we identify a common reference scale by focusing on pure-noise segments. For such segments, the posterior broadly follows the prior (as discussed in Sec.~\ref{sec:bias_origin}), and the information gain is governed primarily by the fraction of the prior supporting signals strong enough to be excluded by the data. The bin priors designed in Sec.~\ref{subsec:setup} are characterized by the range of optimal SNR they support, and we select the prior with the highest SNR bin in our model ($6.5 \leq \rho_{\rm{opt}} < 8$), whose representative optimal SNR is closest to the median SNR scale of the prior used in Ref.~\cite{Kou:2025bhk}. 
We then determine an effective $n_{\rm{live}}$ by requiring that $\sigma_i = \sqrt{H_i/n_{\rm{live}}}$, evaluated on noise segments in this bin prior, matches the average $\sigma_i$ measured for noise segments in the previous analysis.

This procedure maps the uncertainty from a full-dimensional nested sampling run onto our reduced model by matching the information scale of the prior. The resulting $n_{\rm{live}}$ should be interpreted as an effective parameter that reproduces realistic uncertainty levels rather than a literal count of live points. We use this calibrated value uniformly across all bin priors, since a real analysis applies a single sampler configuration to every segment and the variation of $\sigma_i$ between different search prior regions then arises through $H_i$ alone. We emphasize that this calibration likely underestimates the effect: a full 15-dimensional analysis would generally yield larger $H_i$ and consequently larger $\sigma_i$ at the same $n_{\rm{live}}$. The bias observed in our study therefore represents a lower bound on what would be encountered in a realistic analysis.


\begin{figure}[t]
    \centering
    \includegraphics[width=1.0\linewidth]{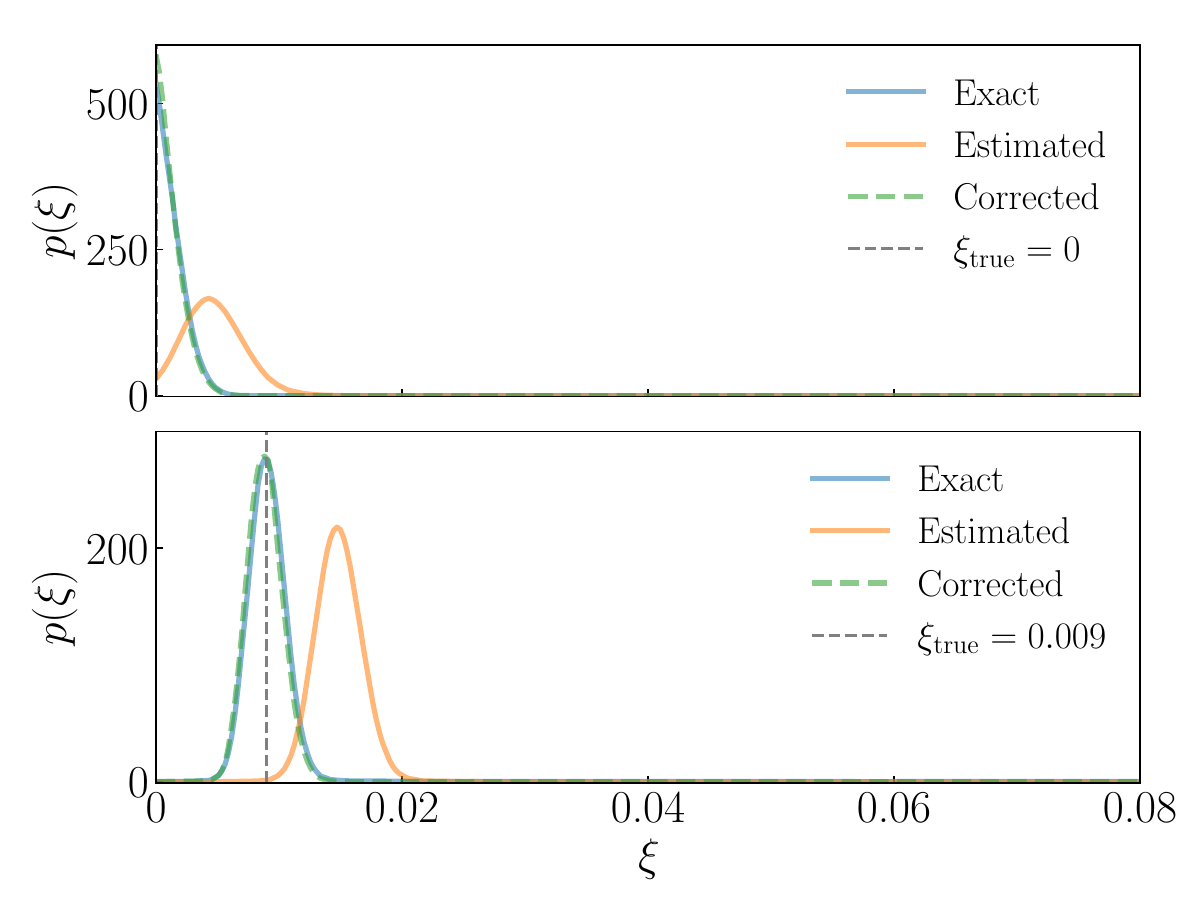}
    \caption{Posterior distributions of the duty cycle $\xi$ from the TBS mock data analysis. The top panel shows the pure-noise case ($\xi_{\rm{true}}=0$) analyzed with 500,000 segments, and the bottom panel shows a dataset with simulated signals at $\xi_{\rm{true}}=0.009$ (vertical dashed line) analyzed with 1,000,000 segments. In each panel, three cases are compared: the posterior obtained using the exact, numerically computed log Bayes factors (blue curves); the posterior using log Bayes factors perturbed by estimation uncertainty (orange curves); and the posterior with the $\frac{1}{2}\sigma^2_i$ correction of Eq.~\ref{eq:corrected_bf} applied (green curves). Ignoring the evidence uncertainty leads to a substantial upward bias in the inferred duty cycle in both cases, while the correction recovers the exact posterior.}
    \label{fig:full_snr_injected}
\end{figure}

Fig.~\ref{fig:full_snr_injected} presents the results of the TBS mock data analysis. The top panel shows the pure-noise case ($\xi_{\rm{true}}=0$), analyzed with 500,000 data segments. Even in the absence of any simulated signal, the uncorrected analysis (orange curve) spuriously infers a nonzero duty cycle, with the posterior peaking near $\xi \sim 0.01$. The posterior after applying correction to each segment closely follows the exact result, accurately recovering a distribution that peaks at zero. The bottom panel shows a dataset of one million segments with a simulated duty cycle of $\xi_{\rm{true}}=0.009$. The uncorrected posterior is again biased toward higher value of $\xi$, while the corrected posterior recovers the simulated value in good agreement with the exact result. These findings confirm that the bias identified in the univariate-Gaussian toy model persists in a frequency-domain TBS analysis with an astrophysically motivated signal population, and that the simple correction of Eq.~\ref{eq:corrected_bf} remains effective.

\subsection{Implications for Searches Targeting Weaker Signals}
\label{subsec:weak_signals}
For source classes such as binary neutron star (BNS) mergers, which are also a potential target~\cite{Smith:2017vfk,HernandezVivanco:2019fku} for TBS, the evidence uncertainty problem is expected to be considerably more severe than for BBH systems. To maintain a well-defined duty cycle definition, the segment duration must be kept short (typically several seconds), since longer segments would raise the probability of multiple overlapping mergers within a single segment~\cite{Johnson:2024foj} and violate the assumption of Eq.~\ref{eq:simple_mixture} that each segment contains at most one signal. Such segments capture primarily the late inspiral and merger phase of the signal. For BBH systems, this phase concentrates a substantial fraction of the total SNR. For BNS mergers, however, the majority of the SNR accumulates over a much longer inspiral lasting minutes to hours in the detector band, and only a small fraction falls within any single seconds long segment. As a result, the per-segment SNR for BNS signals is significantly lower than for BBH signals, placing the signal population deep in much lower optimal SNR bins where a more accurate estimate of evidence per segment may become more important. To investigate this regime, we repeat the TBS mock data analysis restricting the signal population to priors of optimal SNR bins below 4, mimicking a scenario in which the loudest signals available to the search are significantly weaker than in the BBH case. 
\begin{figure}[t]
    \centering
    \includegraphics[width=1.0\linewidth]{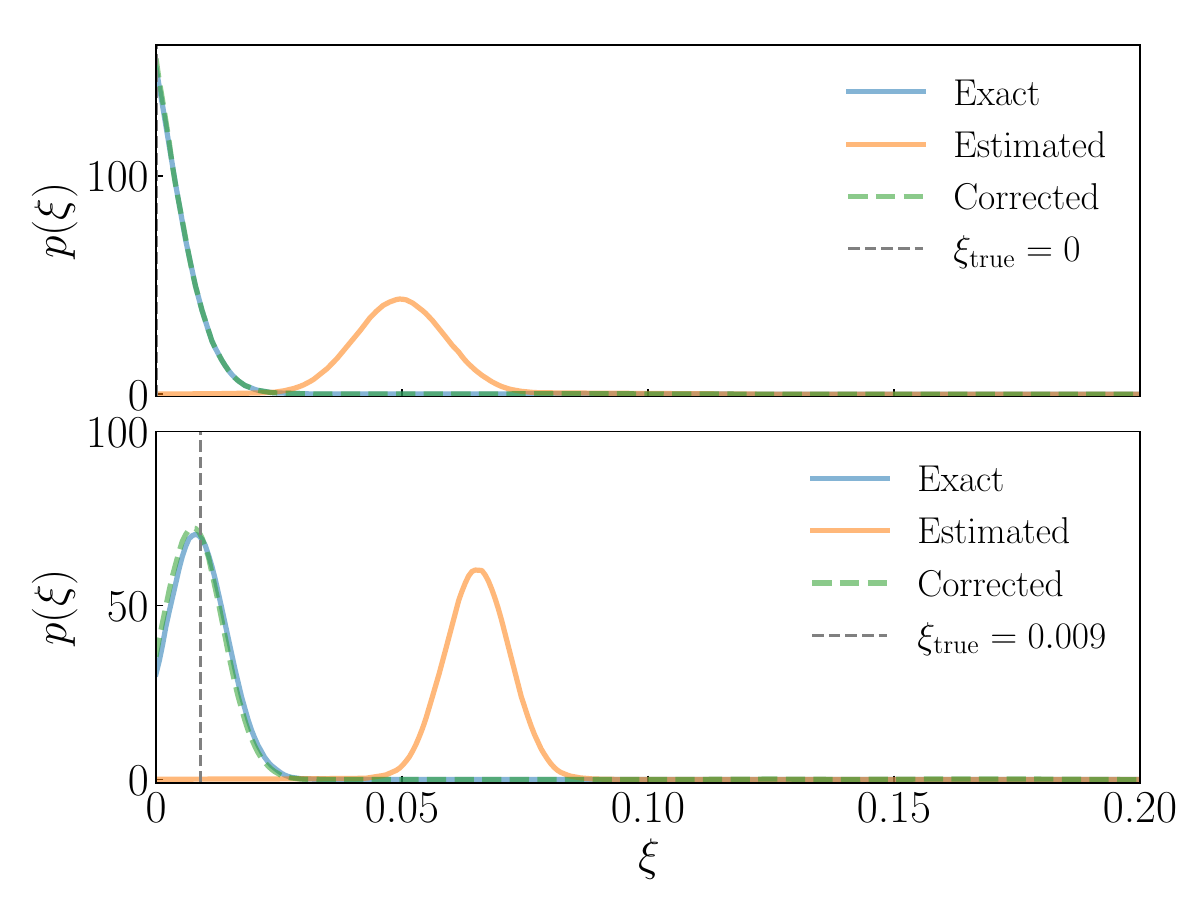}
    \caption{Same as Fig.~\ref{fig:full_snr_injected}, but restricting the signal population and search priors to optimal SNR bins below 4, representative of a scenario dominated by weak signals such as BNS mergers. The top panel shows the pure-noise case $\xi_{\rm{true}}=0$ and the bottom panel a simulated duty cycle of $\xi_{\rm{true}}=0.009$. The uncorrected bias is dramatically amplified compared to the full BBH analysis in Fig.~\ref{fig:full_snr_injected}. The correction of Eq.~\ref{eq:corrected_bf} continues to recover the exact posterior accurately in both cases.}
    \label{fig:low_snr_injected}
\end{figure}

Fig.~\ref{fig:low_snr_injected} shows the results for this reduced-SNR analysis. The top panel presents the pure-noise case ($\xi_{\rm{true}} = 0$) and the bottom panel a simulated duty cycle of $\xi_{\rm{true}}=0.009$. The bias from directly analyzing uncorrected segments is dramatically amplified compared to the full optimal SNR BBH analysis, the posterior analyzed with estimation uncertainty (orange curve) peaks near $\xi \sim 0.06$, roughly seven times the simulated value, and even in the pure-noise case the uncorrected analysis infers a substantial spurious duty cycle. The correction of Eq.~\ref{eq:corrected_bf} continues to recover the exact posterior accurately in both cases. This finding is consistent with the observation in~\cite{HernandezVivanco:2019fku} that BNS searches are more sensitive to systematic effects in the evidence estimation, though that work mitigated the problem by artificially restricting the search prior to low redshifts rather than correcting for the evidence uncertainty directly. The correction proposed here offers a more principled alternative that does not require discarding sensitivity to distant sources.

\begin{figure}[t]
    \centering
    \includegraphics[width=1.0\linewidth]{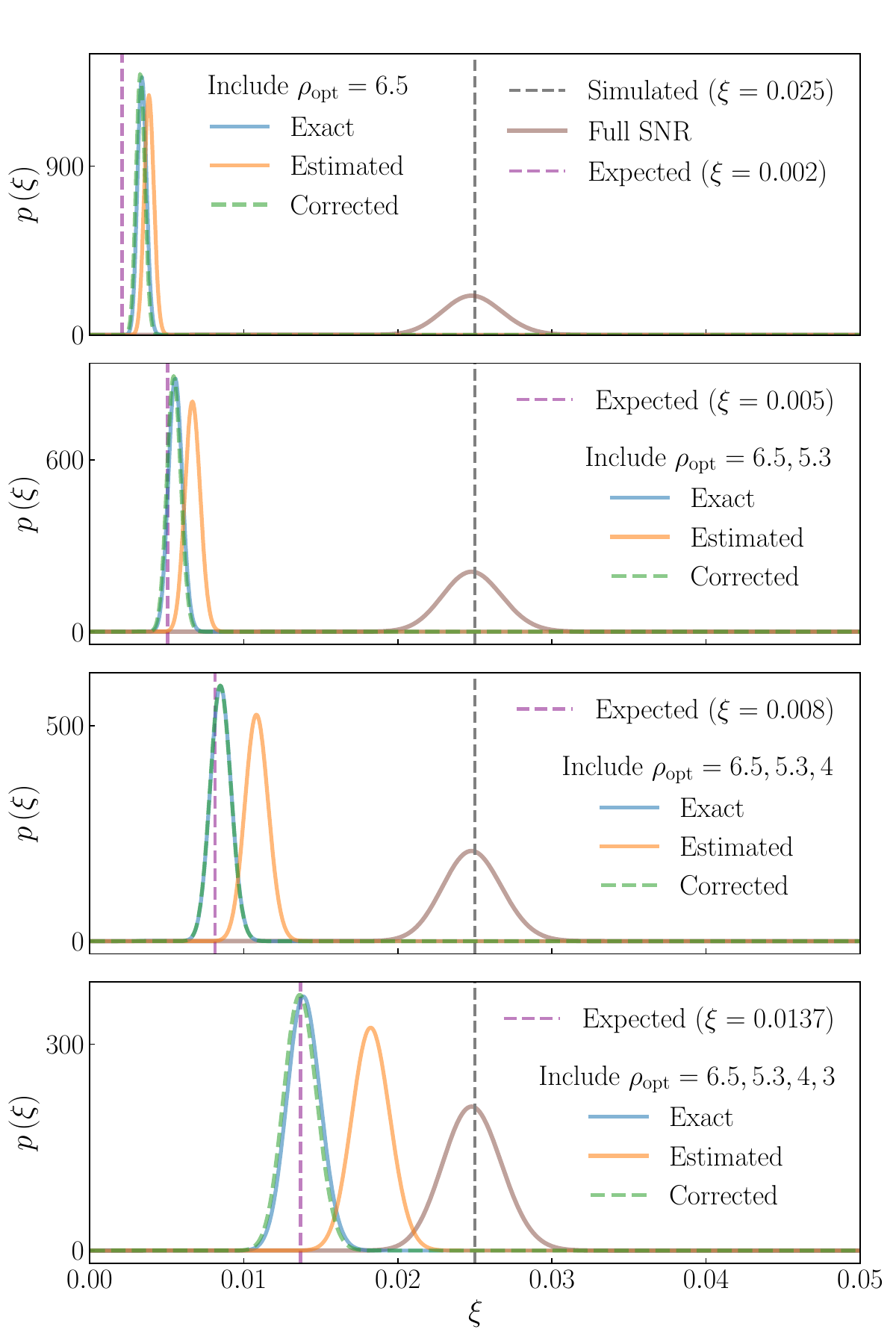}
    \caption{Effect of progressively expanding the search prior on the recovered duty cycle. The data contain one million segments with simulated signals from all optimal SNR bins at a total duty cycle of $\xi_{\rm{true}}=0.025$ (black dashed line), with relative fractions between bins fixed by the population model. Each panel corresponds to a search prior including a different subset of bins: $\rho_{\rm{opt}} = 6.5$ (top), $\rho_{\rm{opt}} = 6.5, 5.3$ (second), $\rho_{\rm{opt}} = 6.5, 5.3$ and $4$ (third), and $\rho_{\rm{opt}} = 6.5, 5.3, 4$ and $3$ (bottom). The purple dashed line indicates the expected duty cycle from signals within the included bins only. Three cases are compared in each panel: the posterior obtained using the exact log Bayes factors (blue), the posterior using log Bayes factors perturbed by estimation uncertainty (orange), and the posterior with the $\frac{1}{2}\sigma_i^2$ correction of Eq.~\ref{eq:corrected_bf} applied (green).The brown curve, repeated in every panel for reference, shows the posterior obtained with the full optimal SNR prior using the exact log Bayes factors.}
    \label{fig:tbs_prior_mismatch}
\end{figure}

Another related question is whether signals at very low optimal SNR contribute meaningfully to the TBS search at all, particularly when the search prior covers only a limited region of parameter space corresponding to louder signals. To investigate this, we construct a mock dataset of one million segments with simulated signals from all prior bins at a total duty cycle of $\xi_{\rm{true}}=0.025$\footnote{We adopt a relatively large total duty cycle for this test to ensure that each SNR bin contains a sufficient number of simulated signals to accurately reflect the population-level strength distribution, even with one million data segments. At smaller duty cycles, the finite number of drawn signals per bin would introduce sampling noise that could obscure the systematic effects we aim to isolate.}, with relative fractions between bins fixed by the population model (see Fig.~\ref{fig:snr_partition}). We then perform the analysis using search priors of increasing breadth, starting from a prior covering only the bin $\rho_{\rm{opt}}=6.5$ and progressively expanding to include $\rho_{\rm{opt}}=5.3,4$ and $3$.

As shown in Fig.~\ref{fig:tbs_prior_mismatch}, when the search prior covers only the loudest bin $\rho_{\rm{opt}}=6.5$, the recovered duty cycle from the exact
log Bayes factors significantly exceeds the expected value $\xi=0.002$ for that bin (the relative error is more than $50\%$). This indicates that signals from unmodeled intermediate SNR bins are being mistakenly absorbed into the modeled bin, inflating the inferred duty cycle. This is not due to the selection effects that appear in population inference~\cite{Mandel:2018mve, Thrane:2018qnx, Vitale:2020aaz}, where a detection threshold determines which events enter the analysis. No segments are excluded and the simulated population is the same in every panel; only the search prior does not match the simulated one. As the search prior is expanded to include $\rho_{\rm{opt}}=5.3$ (second panel) and then $4$ (third panel), the discrepancy between the recovered and expected duty cycles progressively decreases, as the prior now explicitly accounts for a larger fraction of the signals present in the data.  In the bottom panel, where the search prior extends down to $\rho_{\rm{opt}}=3$, the recovered posterior matches the expected duty cycle of $\xi=0.0137$ for the included bins. This agreement indicates that signals with optimal SNR weaker than $\sim3$ are effectively indistinguishable from noise under this search configuration and their presence or absence has no detectable impact on the duty cycle inference, similar to Ref.~\cite{Bers:2025tei}, where weak signals were likewise found to make no detectable contribution to the inferred duty cycle, though using matched-filter SNR as the measure. Signals at intermediate optimal SNR $\rho_{\rm{opt}}=5.3, 4$, by contrast, do contribute to the inferred duty cycle even though they fall outside the nominal search prior: the excess over the expected value in the top two panels shrinks only as those bins are added to the prior. A prior that excludes them therefore does not exclude their influence on the result.


The three cases shown in each panel of Fig.~\ref{fig:tbs_prior_mismatch} separate the effect of the search prior from that of the evidence uncertainty. In every
configuration the corrected posterior recovers the exact result, confirming that the correction of Eq.~\ref{eq:corrected_bf} remains effective regardless of how
the search prior is chosen. The uncorrected posteriors, by contrast, behave differently along the sequence. When the prior covers only the loudest bin, $\rho_{\rm{opt}} = 6.5$, the signal and noise distributions of $\ln \mathrm{BF}$
remain comparatively well separated and the uncorrected posterior departs only mildly from the exact one. As weaker signals are admitted through the prior, the
two distributions start to overlap and the departure grows, mirroring the contrast between Figs.~\ref{fig:full_snr_injected} and \ref{fig:low_snr_injected}. The evidence uncertainty therefore matters more precisely when the prior is broadened to capture the weak-signal population, which is the regime a realistic search has to cover.

This also explains why the bias was not apparent in the analysis of Ref.~\cite{Kou:2025bhk}. The search prior adopted there corresponds closely to our highest optimal SNR bin, the configuration of the top panel in Fig.~\ref{fig:tbs_prior_mismatch}, where the estimation uncertainty has little effect. That analysis also used several thousand segments rather than $\sim 10^6$, too few for the per-segment shift to accumulate into a visible bias. We have checked that applying the correction of Eq.~\ref{eq:corrected_bf} to those results leaves them unchanged.

These two findings reveal a tension in the design of the search prior. Extending the prior to lower optimal SNR regions when they become dominated in the search prior, increases the fraction of weak signals to target for larger detectable total duty cycle but the evidence uncertainty from the sampling becomes problematic, exacerbating the bias if left uncorrected, as demonstrated in Fig.~\ref{fig:low_snr_injected}. Restricting the prior too aggressively\footnote{To our knowledge, existing TBS analyses have not yet employed a comprehensive astrophysical population model; the signal priors adopted in previous studies~\cite{Smith:2017vfk, Smith:2020lkj, Biscoveanu:2020gds, Bers:2025tei} typically support or are dominated by signals with relatively large optimal SNR $\rho_{\rm{opt}}\gtrsim 5$. By contrast, population models extending to high redshifts (e.g., $z\sim 10$) are routinely used in stochastic cross-correlation searches to predict the gravitational-wave energy density spectrum. The evidence uncertainty bias and prior design considerations discussed here become increasingly relevant to the physical interpretation of duty cycle and population inference results.}, on the other hand, leaves the evidence barely affected by estimation uncertainty, but risks misattributing intermediate-SNR signals to the modeled bins, overestimating the duty cycle they can probe in a different way, as shown in Fig.~\ref{fig:tbs_prior_mismatch}. The correction of Eq.~\ref{eq:corrected_bf} resolves one side of this trade-off,
since it allows the prior to be extended to weaker signals without incurring the evidence uncertainty bias. How far it is worth extending the prior, however, remains an open question, since signals with $\rho_{\rm{opt}}\lesssim 3$ seem to leave no imprint on the duty cycle at the one million segments analyzed here. But this does not mean that low SNR signals carry no information: their contribution may simply fall below the statistical uncertainty on $\xi$ at this data volume and prior configuration. How that uncertainty scales with the number of segments as a function of $\rho_{\rm opt}$, and how many segments would therefore be required to resolve the weak-signal contribution, is left for future work.



\section{Discussion and Summary}
\label{sec:summary}
In this work, we have investigated how the statistical uncertainty in evidence estimate per segment propagates into systematic biases in the inferred duty cycle within the Bayesian templated background search framework. Through a univariate-Gaussian toy model and a frequency-domain mock data analysis with an astrophysically motivated BBH population, we have demonstrated that this bias is not a minor technical nuisance but a leading-order systematic effect that can shift the recovered duty cycle by an order of magnitude or more when millions of data segments are accumulated.

The origin of the bias is straightforward: the log Bayes factor for each segment estimated by nested sampling carries uncertainty due to the intrinsic properties of the algorithm, and exponentiating a noisy log-quantity produces an expectation value that is systematically higher than the true value. When this upward shift is compounded across $\sim 10^6$ segments in the mixture likelihood, it coherently inflates the inferred duty cycle. The correction we propose, subtracting $\frac{1}{2}\sigma^2_i$ from each estimated $\log \rm{BF}_i$, before it enters the mixture model (Eq.~\ref{eq:corrected_bf}), removes this leading-order bias by construction. As shown in Appendix~\ref{app:derivation}, the residual bias after correction is suppressed by a factor of $\xi^*$, which in the TBS context is $\leq 10^{-2}$, rendering it negligible.

Our correction relies on the assumption that the log-evidence error is well described by a Gaussian distribution with standard deviation $\sigma_i \sim \sqrt{H_i/n_{\rm{live}}}$. This approximation has been validated both theoretically~\cite{Skilling:2006gxv,Higson2017SamplingEI,Higson:2018cqj,Speagle:2019ivv,Fowlie:2022jls} and numerically~\cite{Fowlie:2022jls} for standard nested sampling implementations. However, it may break down in pathological cases such as strongly multimodal posteriors\footnote{We note that for TBS analyses, the vast majority of data segments are either pure noise or contain very weak signals. For pure noise segments, the posterior remains close to the prior, the information gain $H_i$ is small, and the prior volume compression in nested sampling proceeds smoothly without encountering the multimodal or highly concentrated structures that can cause the Gaussian approximation to break down. The multimodal behavior mentioned here is therefore unlikely to be relevant in practice, as segments containing loud signals with complex posterior structure constitute a negligible fraction of the total dataset.} or extremely low $n_{\rm{live}}$, where the true error distribution could become skewed or heavy-tailed. In such scenarios, the $\frac{1}{2}\sigma^2_i$ correction would no longer reliably remove the leading-order bias. For practical TBS analyses, where $n_{\rm{live}}$ is typically $\sim 1000\text{--}2500$, we expect the Gaussian approximation to be reliable (to the leading order of the potential distribution). Moreover, as we demonstrate in Appendix~\ref{app:validation}, the bias persists even at $n_{\rm live}=3000$, which represents a computationally prohibitive choice for real analyses, confirming that increasing $n_{\rm live}$ alone is not a viable alternative to the correction. The uncertainty $\sigma_i$ captures only the dominant statistical error arising from the stochastic compression of prior volume at each nested sampling iteration. In practice, nested sampling implementations~\cite{Speagle:2019ivv,Buchner:2014nha, Williams:2021qyt,Handley:2015vkr} incur an additional truncation error controlled by the stopping criterion \texttt{dlogz} (GW analyses typically adopt \texttt{dlogz}$\,=0.1$), which terminates the run when the estimated remaining log-evidence contribution falls below this threshold. The correction proposed here addresses only the statistical component. Nevertheless, future work should investigate both the sensitivity of the correction to deviations from Gaussianity and the truncation error using a heavy-tailed model~\cite{Sasli:2023mxr, Karnesis:2024pxh, Sasli:2026pds}.

The analysis of progressively expanding the search prior in Sec.~\ref{subsec:weak_signals} reveals two competing requirements in the design of TBS pipelines. A search prior restricted to high optimal SNR reduces the impact of evidence uncertainty, but causes intermediate-SNR signals to be misattributed to the modeled bins, inflating the duty cycle. 
Extending the prior to lower optimal SNR region removes this misattribution, but the estimation uncertainty then grows and biases the duty cycle if left uncorrected. The search prior has to extend low enough to capture intermediate-strength signals in real searches, and the correction of Eq.~\ref{eq:corrected_bf} is what makes this possible without incurring the accompanying evidence uncertainty bias and becomes increasingly important as the prior is broadened.

Although the primary focus of this work has been the duty cycle, the evidence uncertainty bias inevitably propagates into the joint inference of the BBH population hyperparameters when the mixture model is extended to include them, as discussed in Appendix~\ref{app:pop_bias}. Signals falling outside the search prior could induce a bias in the jointly inferred population parameters along with the duty cycle.
A full study of this effect on joint population inference with realistic BBH population models is warranted and will be pursued in future work.

Based on our findings, we conclude with several practical recommendations for TBS analyses. First, both the log Bayes factor $\ln \rm{BF}_i$ and its associated uncertainty $\sigma_i$ should be stored for every data segment during the search stage; $\sigma_i$ is a standard output of nested sampling packages and incurs negligible additional computational cost. Second, the correction of Eq.~\ref{eq:corrected_bf} should be applied before combining segment-level evidences in the mixture model, which amounts to a one-line modification in existing pipelines~\cite{Kou:2025bhk} and is far more effective than increasing $n_{\rm live}$ (see Appendix~\ref{app:validation}). Finally, the correction becomes indispensable as the search prior is extended to weaker signals, and in particular for source populations dominated by low optimal SNRs, such as BNS mergers, where the SNR is spread over an inspiral far longer than a single segment, and BBH mergers at high redshift, where the signals are intrinsically faint.


\begin{acknowledgments}
The authors acknowledge the computational resources provided by the Minnesota Supercomputing Institute (MSI), which were critical for this study. The authors benefited from discussions with Galin Jones, Rui Zhou, Colm Talbot and Michael Coughlin. We thank Sylvia Biscoveanu for conducting the internal collaboration review that helped improve the quality of the manuscript. We are grateful for computational resources provided by the LIGO Laboratory and supported by National Science Foundation Grants PHY-0757058 and PHY-0823459. X.K, A.S and V.M are supported in part by the NSF grant PHY-2409173. M.S.\ acknowledges the support from Weinberg Institute for Theoretical Physics at the University of Texas at Austin. X.K. was partially supported by the University of Minnesota Data Science Initiative with funding made available by the MnDrive initiative.
This material is based upon work supported by NSF's LIGO Laboratory which is a major facility fully funded by the National Science Foundation.
\end{acknowledgments}

\appendix
\section{Details of Univariate-Gaussian Model}
\label{app:toy_details}
In this appendix, we collect the analytical expressions for the signal and noise evidences, the log Bayes factor, and the information gain used in the univariate-Gaussian toy model of Sec.~\ref{sec:toy_model}.

\subsection{Evidences and Log Bayes Factor}
Under the noise hypothesis, each data point $d_i$ is drawn from a zero-mean Gaussian,
\begin{equation}
\label{eq:gaussian_noise_evidence}
\mathcal{Z}_n(d_i) = \frac{1}{\sqrt{2\pi\sigma_n^2}}\exp\!\left(-\frac{d_i^2}{2\sigma_n^2}\right).
\end{equation}
Under the signal hypothesis, the signal amplitude $x_s$ is drawn from the prior $x_s \sim \mathcal{N}(\mu_s, \sigma^2_s)$ and the observation is $d_i \sim \mathcal{N}(x_s, \sigma^2_n)$. Marginalizing over $x_s$ yields
\begin{equation}
\label{eq:gaussian_signal_evidence}
\begin{aligned}
\mathcal{Z}_s(d_i) &= \int \mathcal{N}(d_i;\, x_s,\, \sigma_n^2)\,\mathcal{N}(x_s;\, \mu_s,\, \sigma_s^2)\,dx_s \\
&= \frac{1}{\sqrt{2\pi(\sigma_n^2 + \sigma_s^2)}}\exp\!\left(-\frac{(d_i - \mu_s)^2}{2(\sigma_n^2 + \sigma_s^2)}\right),
\end{aligned}
\end{equation}
which follows from the standard convolution of two Gaussians. The log Bayes factor for segment $d_i$ is then
\begin{equation}
\label{eq:toy_lnbf}
\ln\mathrm{BF}(d_i) = \frac{1}{2}\ln\!\left(\frac{\sigma_n^2}{\sigma_n^2 + \sigma_s^2}\right) + \frac{d_i^2}{2\sigma_n^2} - \frac{(d_i - \mu_s)^2}{2(\sigma_n^2 + \sigma_s^2)}.
\end{equation}
Under the noise hypothesis, $d_i \sim \mathcal{N}(0,\sigma_n^2)$, so $\mathbb{E}[d^2_i] = \sigma^2_n$ and
\begin{equation}
\mathbb{E}[\ln\mathrm{BF}_i\,|\, N] = \frac{1}{2}\ln\!\left(\frac{\sigma_n^2}{\sigma_n^2 + \sigma_s^2}\right) + \frac{\sigma_s^2 - \mu^2_s}{2(\sigma_n^2 + \sigma_s^2)}.
\end{equation}
Under the signal-plus-noise hypothesis, $d_i \sim \mathcal{N}(\mu_s, \sigma^2_n + \sigma^2_s)$ so $\mathbb{E}[d^2_i]=\mu^2_s + \sigma^2_n + \sigma^2_s$ and
\begin{equation}
\mathbb{E}[\ln\mathrm{BF}_i\,|\,S] = \frac{1}{2}\ln\!\left(\frac{\sigma_n^2}{\sigma_n^2 + \sigma_s^2}\right) + \frac{\mu^2_s + \sigma^2_s}{2\sigma^2_n}.
\end{equation}
For our parameter choices ($\mu_s = 0, \sigma_s=2, \sigma_n=3$), we find $\mathbb{E}[\ln\mathrm{BF}_i\,|\, N] \approx -0.03$ and $\mathbb{E}[\ln\mathrm{BF}_i\,|\, S] \approx 0.04$.
\subsection{Information Gain}
The posterior on the signal amplitude $x_s$ given observation $d_i$ under the signal hypothesis is
\begin{equation}
p(x_s\,|\,d_i,\, S) = \mathcal{N}(x_s;\,\mu_{\rm post},\,\sigma_{\rm post}^2),
\end{equation}
where
\begin{equation}
\sigma_{\rm post}^2 = \frac{\sigma_s^2\,\sigma_n^2}{\sigma_s^2 + \sigma_n^2}, \quad \mu_{\rm post} = \sigma_{\rm post}^2\left(\frac{\mu_s}{\sigma_s^2} + \frac{d_i}{\sigma_n^2}\right).
\end{equation}
For $\mu_s=0$, this reduces to $\mu_{\rm{post}}=\sigma^2_{\rm post}d_i/\sigma^2_n$. The information gain (KL divergence from prior to posterior) for segment $d_i$ is
\begin{equation}
\label{eq:toy_KL}
\begin{aligned}
H_i &= D_{\rm KL}\!\left[p(x_s|d_i, S)\,\|\,\pi(x_s)\right] \\ 
&= \frac{1}{2}\ln\!\left(\frac{\sigma_s^2}{\sigma_{\rm post}^2}\right) + \frac{\sigma_{\rm post}^2 + (\mu_{\rm post} - \mu_s)^2}{2\sigma_s^2} - \frac{1}{2},
\end{aligned}
\end{equation}
which follows from the standard KL divergence between two univariate Gaussians.

\section{Biases on Population Inference}
\label{app:pop_bias}
\begin{figure*}[ht]
    \centering
    \includegraphics[width=0.49\linewidth]{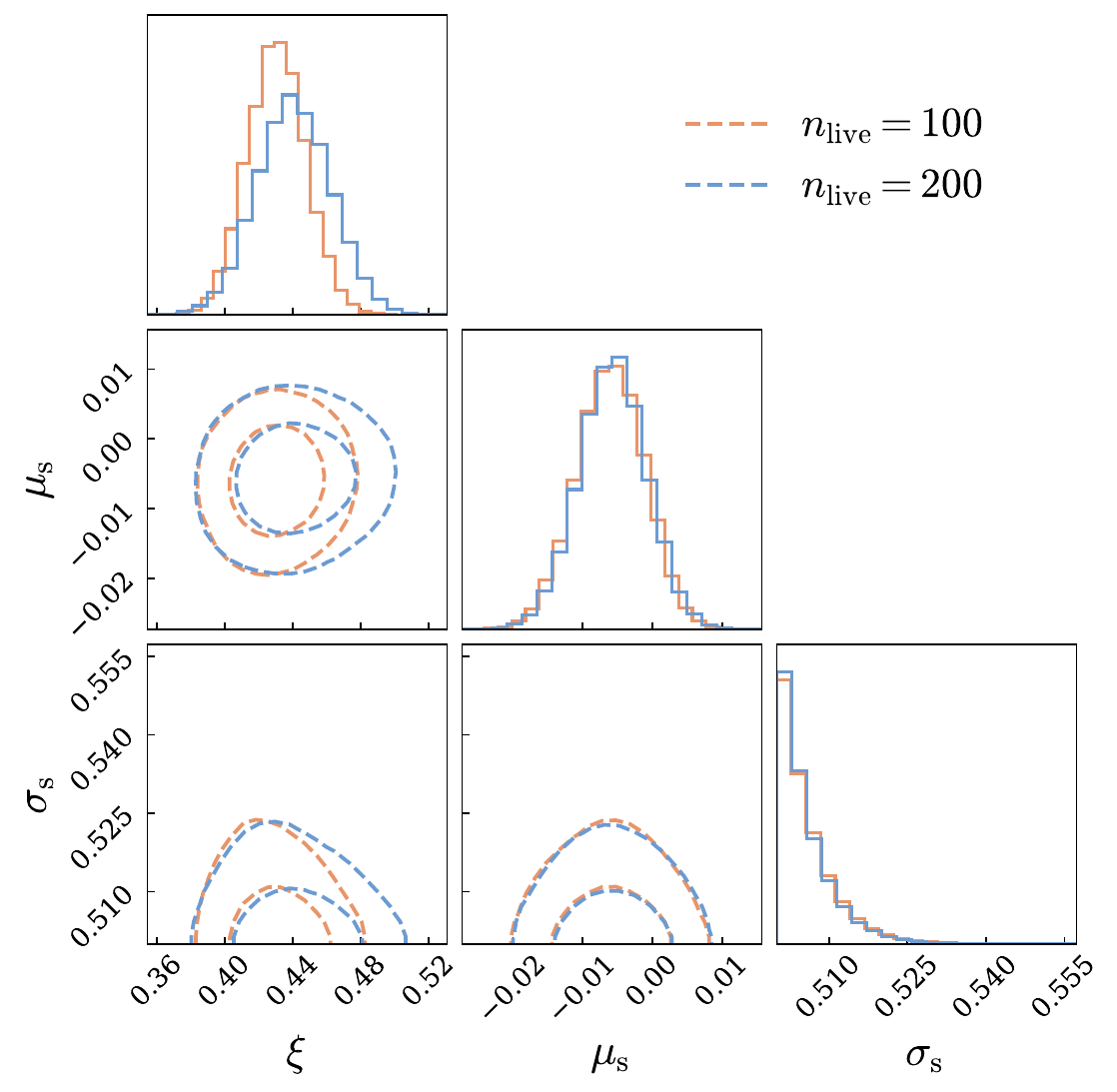}
    \includegraphics[width=0.49\linewidth]{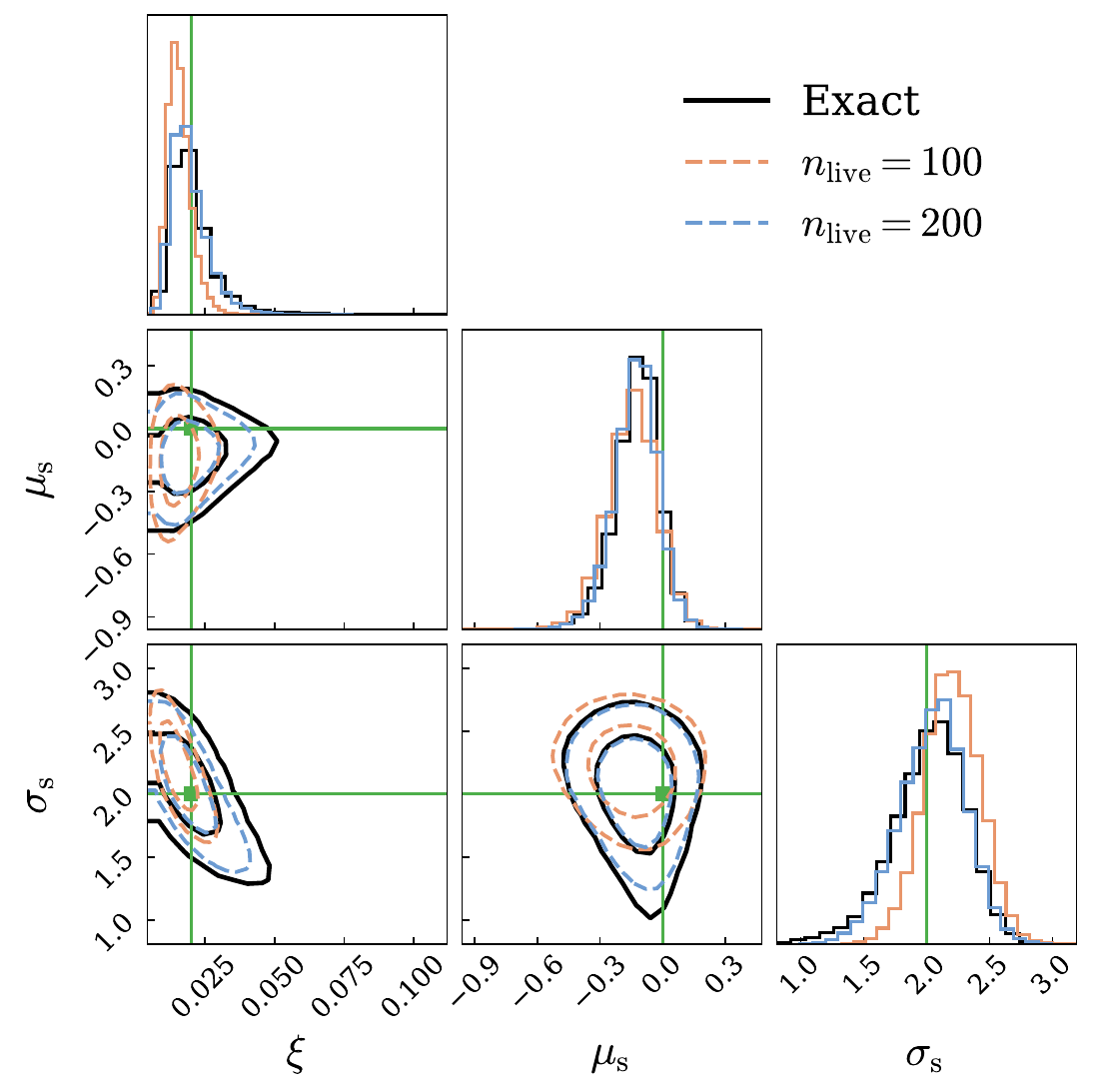}
    \caption{Impact of evidence uncertainty on joint population inference in the univariate-Gaussian toy model with $\xi_{\rm{true}}=0.02, \mu_{s,\rm{true}}=0$ and $\sigma_{s,\rm{true}}=2$. Left: joint posterior on $(\xi, \mu_s, \sigma_s)$ obtained without the correction, using log Bayes factors perturbed at $n_{\rm{live}} = 100$ (orange) and $200$ (blue). Right: the same analysis with the correction of Eq.~\ref{eq:corrected_bf} applied, together with the exact result (black); green lines mark the simulated values. The corrected posterior recovers the simulated values in excellent agreement with the exact result.}
    \label{fig:pop_corner}
\end{figure*}
The analysis in Sec.~\ref{sec:toy_model} and ~\ref{sec:tbs_mock} focused on the inference of the duty cycle $\xi$ with the population parameters held fixed at their true values. In practice, however, the TBS framework is designed to jointly infer $\xi$ together with the hyperparameters governing the signal population~\cite{Smith:2017vfk, Smith:2020lkj, Biscoveanu:2020gds, Bers:2025tei}. In this Appendix, we extend the univariate-Gaussian toy model to this joint inference setting and demonstrate that uncorrected evidence uncertainty biases not only $\xi
$ but also the recovered population hyperparameters.

We promote the signal distribution parameters $\mu_s$ and $\sigma_s$ to free parameters and perform a joint inference of $(\xi, \mu_s, \sigma_s)$ using the mixture model of Eq.~\ref{eq:total_mixture}, where the signal evidence $\mathcal{Z}^i_s$ now depends on $\mu_s$ and $\sigma_s$ through Eq.~\ref{eq:gaussian_signal_evidence} while the noise evidence $\mathcal{Z}^i_n$ is independent of the hyperparameters. The hierarchical likelihood is then
\begin{equation}
\mathcal{L}(D \,|\, \xi, \mu_s, \sigma_s) = \prod_i^N
\left[\xi\, \mathcal{Z}^i_s(\mu_s, \sigma_s) + (1-\xi)\, \mathcal{Z}^i_n\right],
\end{equation}
which we sample with \textsc{emcee}~\cite{Foreman-Mackey:2012any} under uniform priors $\xi \in [0, 1]$, $\mu_s \in [-10, 10]$ and $\sigma_s \in [0.5, 15]$. Unlike
typical GW population inference, no reweighting of per-event posterior samples is required here: both evidences are available in closed form, so
$\mathcal{Z}^i_s$ can be re-evaluated directly at every hyperparameter draw.

We generate a mock dataset of two million segments with simulated values $\xi_{\rm{true}}=0.02$, $\mu_{s,\rm{true}}=0$ and $\sigma_{s,\rm{true}}=2$, and analyze it under three conditions: using the exact log Bayes factors, using log Bayes factors perturbed with $\sigma_i = \sqrt{H_i/n_{\rm{live}}}$ for $n_{\rm{live}}=100$ and $200$ without correction, and with the correction of Eq.~\ref{eq:corrected_bf} applied. The per-segment perturbations are generated once and held fixed throughout the inference. We compute $H_i$ for every segment at the fiducial values $\mu_s=0$ and $\sigma_s=2$, draw a single $\epsilon_i \sim \mathcal{N}(0, \sigma^2_i)$ per segment with 
$\sigma_i = \sqrt{H_i/n_{\rm live}}$, and apply it to the log Bayes factor evaluated at each hyperparameter draw. This mirrors a real analysis, in which nested sampling is run once per segment at the search stage and the resulting 
$\ln \tilde{\rm{BF}}_i$ and $\sigma_i$ are stored and reused when the mixture model is subsequently evaluated over hyperparameters.

The left panel of Fig.~\ref{fig:pop_corner} shows the uncorrected result. The posterior of $\xi$ peaks around $0.4$, more than an order of magnitude above the simulated value and a far more severe bias than in the fixed-population analysis of Sec.~\ref{sec:toy_model}. This amplification occurs because $\mu_s$ and $\sigma_s$ are now free to adjust: the inference compensates for the spurious excess of apparent signals by driving $\sigma_s$ down against its prior boundary, narrowing the signal distribution to absorb the bias. The right panel shows the exact and corrected cases. The corrected posterior recovers all three simulated values in agreement with the exact result at both $n_{\rm live} = 100$ and $200$, confirming the effectiveness of Eq.~\ref{eq:corrected_bf} in the context of joint population inference. A full investigation using realistic, multi-dimensional BBH population models is beyond the scope of this work and will be pursued in a future study.

\section{Derivation of the Bias and Its Correction}
\label{app:derivation}
In this appendix, we derive the leading-order bias in the inferred duty cycle $\xi$ introduced by evidence estimation uncertainty, and show that the simple correction of Eq.~\ref{eq:corrected_bf} suppresses this bias by a factor of $\xi^*$.

Our starting point is the total log-likelihood for the duty cycle, obtained by taking the logarithm of Eq.~\ref{eq:total_mixture},
\begin{equation}
\ln\mathcal{L}(\xi) = \sum_{i=1}^N \ln\!\left[\xi\,\mathrm{BF}_i + (1-\xi)\right],
\end{equation}
where we have divided each term by $\mathcal{Z}^i_n$ so that the argument of the logarithm is expressed in terms of the Bayes factor $\rm{BF}_i = \mathcal{Z}^i_s/\mathcal{Z}^i_n$. This discards an additive constant $\sum_i \ln \mathcal{Z}^i_n$, which is independent of $\xi$ and therefore does not affect the likelihood. The score function and the observed Fisher information are
\begin{equation}
\begin{aligned}
S(\xi) &= \frac{\partial\ln\mathcal{L}}{\partial\xi} = \sum_{i=1}^N T_i(\mathrm{BF}_i), \\
I(\xi) &= -\frac{\partial^2\ln\mathcal{L}}{\partial\xi^2} = \sum_{i=1}^N \frac{(\mathrm{BF}_i - 1)^2}{A_i(\xi)^2},
\end{aligned}
\end{equation}
where the per-segment contribution to the score is
\begin{equation}
T_i(\mathrm{BF}_i) = \frac{\mathrm{BF}_i - 1}{A_i(\xi)}, \qquad A_i(\xi) \equiv \xi\,\mathrm{BF}_i + (1-\xi),
\end{equation}
with derivatives
\begin{equation}
\label{eq:dT_dBF}
\frac{dT_i}{d\mathrm{BF}_i} = \frac{1}{A_i(\xi)^2}, \quad \frac{d^2T_i}{d\mathrm{BF}_i^2} = -\frac{2\xi}{A_i(\xi)^3}.
\end{equation}
As established in Sec.~\ref{subsec:source_bias}, the nested sampling estimate of $\ln \rm{BF}_i$ carries a Gaussian residual $\epsilon_i \sim \mathcal{N}(0, \sigma_i^2)$ relative to the true value $\ln \mathrm{BF}_i^*$. We define the uncorrected and corrected Bayes factors as
\begin{equation}
\begin{aligned}
\tilde{\mathrm{BF}}_i &= \mathrm{BF}_i^*\,\tilde{U}_i, \quad &\tilde{U}_i &= \exp(\epsilon_i), \\
\hat{\mathrm{BF}}_i &= \mathrm{BF}_i^*\,\hat{U}_i, \quad &\hat{U}_i &= \exp\!\left(\epsilon_i - \tfrac{1}{2}\sigma_i^2\right),
\end{aligned}
\end{equation} 
such that $\mathbb{E}[\tilde{U}_i]=\exp{(\sigma^2_i/2)}$ while $\mathbb{E}[\hat{U}_i]=1$ by construction.

Let $\xi^*$ denote the maximum-likelihood estimate obtained with perfect knowledge of all $\rm{BF}^*_i$, i.e., the solution to $S(\xi^*)=0$. When the Bayes factors are replaced by their noisy estimates, the score at $\xi^*$ is no longer zero in general. Expanding the noisy score around $\xi^*$ to first order,
\begin{equation}
0 = S_{\rm noisy}(\xi) \approx S_{\rm noisy}(\xi^*) + \frac{\partial S}{\partial\xi}\bigg|_{\xi^*}(\xi - \xi^*),
\end{equation}
and taking the expectation over the noise realizations in the limit of large $N$, the asymptotic bias is
\begin{equation}
\mathbb{E}\!\left[\xi - \xi^*\right] \approx \frac{\mathbb{E}\!\left[S_{\rm noisy}(\xi^*)\right] - S(\xi^*)}{I(\xi^*)}.
\end{equation}
The bias is therefore determined by the shift in the expected score at the true maximum-likelihood point, normalized by the Fisher information. 

To evaluate $\mathbb{E}\!\left[S_{\rm noisy}(\xi^*)\right]$, we Taylor expand each per-segment score contribution $T_i(\rm{BF}^*_iU_i)$ around $U_i=1$. For the uncorrected case, since $\mathbb{E}[\hat{U}_i]=\exp(\sigma_i^2/2)\neq 1$ , the leading contribution enters at first order,
\begin{equation}
\mathbb{E}\!\left[T_i(\mathrm{BF}_i^*\tilde{U}_i)\right] \approx T_i(\mathrm{BF}_i^*) + \tfrac{1}{2}\sigma_i^2\,\mathrm{BF}_i^*\,\frac{dT_i}{d\mathrm{BF}_i^*}.
\end{equation}
For the corrected case, since $\mathbb{E}[\hat{U}_i]=1$ by construction, the first-order term vanishes and the leading contribution arises at second order through $\rm{Var}(\hat{U}_i) = \exp(\sigma^2_i)-1$,
\begin{equation}
\mathbb{E}\!\left[T_i(\mathrm{BF}_i^*\hat{U}_i)\right] \approx T_i(\mathrm{BF}_i^*) + \tfrac{1}{2}\,\mathrm{Var}(\hat{U}_i)\,\mathrm{BF}_i^{*2}\,\frac{d^2T_i}{d\mathrm{BF}_i^{*2}}.
\end{equation}
Substituting the derivatives and summing over all $N$ segments, the biases for the uncorrected and corrected cases are
\begin{equation}
\begin{aligned}
\mathbb{E}\!\left[\tilde{\xi} - \xi^*\right] &\approx \frac{\displaystyle\sum_{i=1}^N \frac{\frac{1}{2}\sigma_i^2\,\mathrm{BF}_i^*}{A_i(\xi^*)^2}}{\displaystyle\sum_{i=1}^N \frac{(\mathrm{BF}_i^* - 1)^2}{A_i(\xi^*)^2}}, \\[10pt]
\mathbb{E}\!\left[\hat{\xi} - \xi^*\right] &\approx -\frac{\displaystyle\sum_{i=1}^N \frac{\xi^*\,\mathrm{BF}_i^{*2}\left(\exp(\sigma_i^2) - 1\right)}{A_i(\xi^*)^3}}{\displaystyle\sum_{i=1}^N \frac{(\mathrm{BF}_i^* - 1)^2}{A_i(\xi^*)^2}}.
\end{aligned}
\end{equation}
The critical difference is that the numerator of the corrected expression carries an explicit factor of $\xi^*$, originating from the second derivative $\frac{d^2 T_i}{d \rm{BF}^2_i}$, given in Eq.~\ref{eq:dT_dBF}. In the TBS context where $\xi^* \lesssim 10^{-2}$, this additional suppression renders the residual bias after correction negligibly small compared to the uncorrected case, consistent with the numerical results presented in Secs.~\ref{sec:toy_model} and~\ref{sec:tbs_mock}.

\section{BBH Parameters Reduction}
\label{app:tbs_likelihood}
In this appendix, we describe the reduction of the full BBH parameter space to the effective parameters used in the mock data challenge, and derive the resulting expressions for the likelihood, the Bayes factor, and the information gain.

\subsection{Parameter Reduction}
For a single ground-based detector and a non-precessing, equal-mass and dominant-mode waveform such as \textsc{IMRPhenomXAS}~\cite{Pratten:2020fqn}, the detector response to a gravitational-wave signal can be written as
\begin{equation}
\begin{aligned}
h_{\rm det}(f) =& \left[F_+(\alpha,\delta,\psi)\frac{1+\cos^2\iota}{2} - iF_\times(\alpha,\delta,\psi)\cos\iota\right] \\
&\times h_0(f;\,\mathcal{M}_c, d_L),
\end{aligned}
\end{equation}
where $F_{+,\times}$ are the antenna pattern functions depending on the right ascension $\alpha$, declination $\delta$, and polarization angle $\psi$; $\iota$ is the inclination angle, and $h_0$ is the intrinsic waveform depending on the detector-frame chirp mass $\mathcal{M}_c$ and luminosity distance $d_L$. For the aligned-spin, dominant-mode waveform considered here, the complex bracket can be written as a single effective amplitude and phase,
\begin{equation}
h_{\rm det}(f) = \Theta\times\,|h_0(f;\,\mathcal{M}_c, d_L)|\,e^{i(\Psi(f) + \psi_F)},
\end{equation}
where $\Theta \in [0,1]$ is the effective amplitude factor absorbing the antenna pattern functions and inclination, and $\psi_F$ is an effective phase that can be absorbed into the coalescence phase $\phi_c$. We fix $\phi_c=0$ throughout this analysis, so that the signal in each segment is characterized by $(\mathcal{M}_c, d_L, \Theta)$. Within each optimal SNR bin, the luminosity distance is determined by $\mathcal{M}_c$ and the representative $\rho_{\rm opt}$, further reducing the free parameters to $(\mathcal{M}_c, \Theta)$. The distribution of $\Theta$ across isotropically distributed sky locations, polarizations, and inclinations is well approximated by a $\rm Beta(2,4)$ distribution on $[0,1]$~\cite{Fishbach:2019ckx,Farah:2023vsc}.

We note that despite the dimensional reduction, our model faithfully captures the main features of the $\ln \rm{BF}_i$ distribution that would arise in a full parameter estimation. This is because the effective detector response factor $\Theta$, which enters as a single parameter in our reduced likelihood, is a compact representation of the antenna pattern functions, sky location, polarization angle, and source inclination that appear in the full 15-dimensional integral. The marginalization over $\Theta$, therefore, accounts for the same physical modulation of signal amplitude that would be achieved by a full analysis integrating over multiple extrinsic parameters, hence preserving the statistical structure of the Bayes factor distribution across segments.

\subsection{Likelihood, Bayes Factor and Information Gain}
For stationary Gaussian noise, the frequency-domain likelihood for a data segment $d_i$ given a signal template $h$ is the Whittle likelihood,
\begin{equation}
\mathcal{L}(d_i\,|\,h) \propto \exp\!\left[-\frac{1}{2}\langle d_i - h\,|\,d_i - h\rangle\right],
\end{equation}
where the noise-weighted inner product is defined as
\begin{equation}
\langle a\,|\,b \rangle = 4\,\Re\int_0^\infty \frac{a^*(f)\,b(f)}{S_n(f)}\,df,
\end{equation}
with $S_n(f)$ the one-sided noise power spectral density. The log-likelihood ratio between the signal and noise hypotheses is
\begin{equation}
\ln\frac{\mathcal{L}(d_i\,|\,h)}{\mathcal{L}(d_i\,|\,N)} = \langle d_i\,|\,h\rangle - \frac{1}{2}\langle h\,|\,h\rangle.
\end{equation}
Substituting $h = \Theta h_0$ and defining the matched-filter output $x_i(\mathcal{M}_c) = \langle d_i\,|\, h_0(\mathcal{M}_c)\rangle$ and the squared optimal SNR $\rho^2_{\rm{opt}} = \langle h_0\,|\, h_0 \rangle$, this becomes
\begin{equation}
\ln\frac{\mathcal{L}(d_i\,|\,\mathcal{M}_c, \Theta)}{\mathcal{L}(d_i\,|\,N)} = \Theta\,x_i - \frac{1}{2}\Theta^2\,\rho_{\rm opt}^2 \equiv E(\Theta;\,\mathcal{M}_c).
\end{equation}
The Bayes factor for segment $d_i$ is obtained by marginalizing over both $\mathcal{M}_c$ and $\Theta$.  Since the integrand factorizes, the marginalization over $\Theta$ at fixed $\mathcal{M}_c$ can be performed first,
\begin{equation}
\begin{aligned}
\mathrm{BF}(& x_i,\,\rho_{\rm opt}) = \int_0^1\exp(E(\Theta;\,\mathcal{M}_c))\pi(\Theta)\,d\Theta\\
&=\int_0^1 \exp\!\left(\Theta\,x_i - \frac{1}{2}\rho_{\rm opt}^2\,\Theta^2\right)\,20\,\Theta\,(1-\Theta)^3\,d\Theta,
\end{aligned}
\end{equation}
where the term $20\,\Theta\,(1-\Theta)^3$ comes from the probability density of $\rm Beta(2,4)$ distribution. The total Bayes factor is then obtained by integrating over the chirp mass distribution $\pi(\mathcal{M}_c)$ within each optimal SNR bin,
\begin{equation}
\mathrm{BF}_i = \int \mathrm{BF}(x_i(\mathcal{M}_c),\,\rho_{\rm opt})\,\pi(\mathcal{M}_c)\,d\mathcal{M}_c.
\end{equation}
Since the integrand of $\mathrm{BF}(x_i,\,\rho_{\rm opt})$ is a product of a Gaussian exponential in $\Theta$ and a polynomial, the integral over $\Theta$ can be evaluated analytically in terms of the error function and exponentials via a Gaussian moment recurrence relation. The outer integral over $\mathcal{M}_c$ is evaluated by direct numerical quadrature.

The information gain for each segment can likewise be obtained in closed form that involves the full joint posterior over both $\Theta$ and $\mathcal{M}_c$. By Bayes' theorem, $p(\Theta, \mathcal{M}_c\,|\, d_i) = \exp(E(\Theta;\,\mathcal{M}_c))\pi(\Theta)\pi(\mathcal{M}_c)/\mathrm{BF}_i$, so
\begin{equation}
\begin{aligned}
H_i &= \int\!\!\int p(\Theta, \mathcal{M}_c\,|\,d_i)\,\ln\frac{p(\Theta, \mathcal{M}_c\,|\,d_i)}{\pi(\Theta)\,\pi(\mathcal{M}_c)}\,d\Theta\,d\mathcal{M}_c \\
&= \mathbb{E}_{\rm post}\!\left[E(\Theta;\,\mathcal{M}_c)\right] - \ln\mathrm{BF}_i,
\end{aligned}
\end{equation}
where
\begin{equation}
\begin{aligned}
\mathbb{E}_{\rm post}\left[E(\Theta;\,\mathcal{M}_c)\right] &= \frac{1}{\mathrm{BF}_i}\int\int_0^1 E(\Theta;\,\mathcal{M}_c)\,e^{E(\Theta;\,\mathcal{M}_c)}\\
&\times\,\pi(\Theta)\,d\Theta \pi(\mathcal{M}_c)\,d\mathcal{M}_c.
\end{aligned}
\end{equation}
The inner integral over $\Theta$ at fixed $\mathcal{M}_c$ is again computable analytically through the same moment recurrence extended to higher order, and the outer integral over $\mathcal{M}_c$ is evaluated by numerical quadrature. This two-step procedure, analytical marginalization over $\Theta$ followed by numerical integration over $\mathcal{M}_c$, makes it computationally feasible to obtain ground-truth Bayes factors and information gains for millions of segments.

\section{Validation at Higher $n_{\rm live}$}
\label{app:validation}

A natural question is whether the evidence uncertainty bias can be mitigated simply by increasing the number of live points in the nested sampling run, rather than applying the correction of Eq.~\ref{eq:corrected_bf}. To investigate this, we repeat the TBS mock data analysis with 1,250,000 data segments and per-segment uncertainties scaled to correspond to an effective $n_{\rm live}=3000$. This represents an extreme choice for real data analyses: running nested sampling with $n_{\rm live}=3000$ on each of the millions of segments required for a TBS search would be computationally prohibitive with current resources.

\begin{figure}[t]
    \centering
    \includegraphics[width=1.0\linewidth]{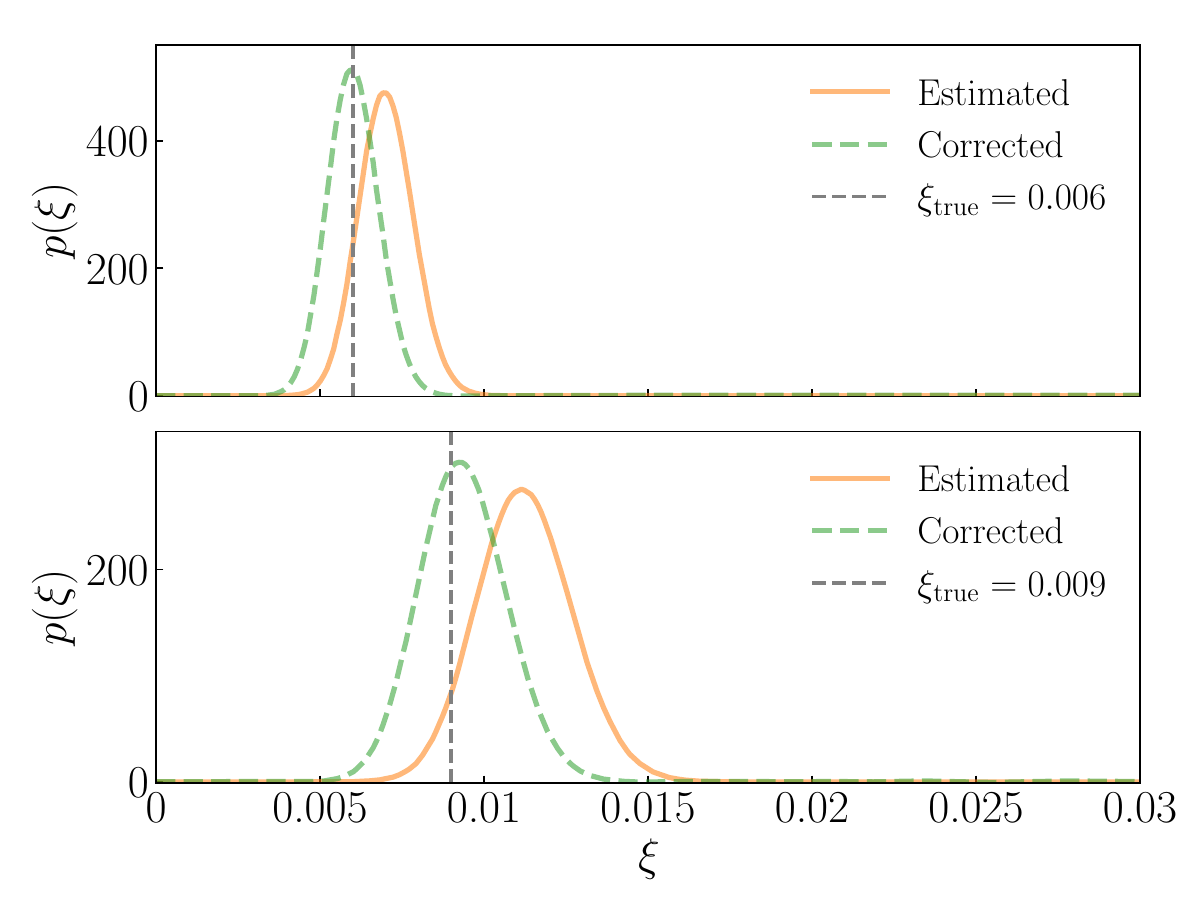}
    \caption{Posterior distributions of the duty cycle $\xi$ from the TBS mock data analysis with 1,250,000 segments and per-segment uncertainties scaled to an effective $n_{\rm live}=3000$. The top panel uses a restricted search prior excluding the $\rho_{\rm opt} < 2$ bins, with a simulated duty cycle of $\xi_{\rm true}=0.006$ (dashed line). The bottom panel uses the full SNR prior, with $\xi_{\rm true}=0.009$. In both cases, the uncorrected posteriors (orange curves) remain visibly shifted from the simulated values, while the corrected posteriors (green curves) recover it accurately.}
    \label{fig:high_nlive}
\end{figure}

Fig.~\ref{fig:high_nlive} shows the results under two prior configurations. The top panel uses a restricted search prior that excludes the lowest optimal SNR bins ($\rho_{\rm opt} < 2$), with a simulated duty cycle of $\xi=0.006$. The bottom panel uses the full SNR prior covering all bins, with a simulated duty cycle of $\xi = 0.009$. In both cases, a visible shift between the estimated and corrected posteriors persists even at $n_{\rm live}=3000$, confirming that the bias is a structural feature of the mixture model that diminishes with increasing $n_{\rm live}$ but does not vanish at practically achievable values. The bias is more pronounced in the bottom panel, where the full prior admits weaker signals with larger relative evidence uncertainty, consistent with the findings of Sec.~\ref{subsec:weak_signals}. These results reinforce our conclusion that the correction of Eq.~\ref{eq:corrected_bf} is the more principled approach: it eliminates the bias regardless of the choice of $n_{\rm live}$ or search prior, without sacrificing sensitivity to any part of the signal population.

\bibliography{refs}

@article{LIGOScientific:2014pky,
    author = "Aasi, J. and others",
    collaboration = "LIGO Scientific",
    title = "{Advanced LIGO}",
    eprint = "1411.4547",
    archivePrefix = "arXiv",
    primaryClass = "gr-qc",
    doi = "10.1088/0264-9381/32/7/074001",
    journal = "Class. Quant. Grav.",
    volume = "32",
    pages = "074001",
    year = "2015"
}

@article{VIRGO:2014yos,
    author = "Acernese, F. and others",
    collaboration = "VIRGO",
    title = "{Advanced Virgo: a second-generation interferometric gravitational wave detector}",
    eprint = "1408.3978",
    archivePrefix = "arXiv",
    primaryClass = "gr-qc",
    doi = "10.1088/0264-9381/32/2/024001",
    journal = "Class. Quant. Grav.",
    volume = "32",
    number = "2",
    pages = "024001",
    year = "2015"
}

@article{KAGRA:2018plz,
    author = "Akutsu, T. and others",
    collaboration = "KAGRA",
    title = "{KAGRA: 2.5 Generation Interferometric Gravitational Wave Detector}",
    eprint = "1811.08079",
    archivePrefix = "arXiv",
    primaryClass = "gr-qc",
    reportNumber = "JGW-P1809243",
    doi = "10.1038/s41550-018-0658-y",
    journal = "Nature Astron.",
    volume = "3",
    number = "1",
    pages = "35--40",
    year = "2019"
}

@article{KAGRA:2013rdx,
    author = "Abbott, B. P. and others",
    collaboration = "KAGRA, LIGO Scientific, Virgo",
    title = "{Prospects for observing and localizing gravitational-wave transients with Advanced LIGO, Advanced Virgo and KAGRA}",
    eprint = "1304.0670",
    archivePrefix = "arXiv",
    primaryClass = "gr-qc",
    reportNumber = "LIGO-P1200087, VIR-0288A-12, JGW-P1808427",
    doi = "10.1007/s41114-020-00026-9",
    journal = "Living Rev. Rel.",
    volume = "19",
    pages = "1",
    year = "2016"
}

@article{LIGOScientific:2017zlf,
    author = "Abbott, Benjamin P. and others",
    collaboration = "LIGO Scientific, Virgo",
    title = "{GW170817: Implications for the Stochastic Gravitational-Wave Background from Compact Binary Coalescences}",
    eprint = "1710.05837",
    archivePrefix = "arXiv",
    primaryClass = "gr-qc",
    reportNumber = "LIGO-P1700272",
    doi = "10.1103/PhysRevLett.120.091101",
    journal = "Phys. Rev. Lett.",
    volume = "120",
    number = "9",
    pages = "091101",
    year = "2018"
}

@article{LIGOScientific:2018mvr,
    author = "Abbott, B. P. and others",
    collaboration = "LIGO Scientific, Virgo",
    title = "{GWTC-1: A Gravitational-Wave Transient Catalog of Compact Binary Mergers Observed by LIGO and Virgo during the First and Second Observing Runs}",
    eprint = "1811.12907",
    archivePrefix = "arXiv",
    primaryClass = "astro-ph.HE",
    reportNumber = "LIGO-P1800307",
    doi = "10.1103/PhysRevX.9.031040",
    journal = "Phys. Rev. X",
    volume = "9",
    number = "3",
    pages = "031040",
    year = "2019"
}

@article{LIGOScientific:2020ibl,
    author = "Abbott, R. and others",
    collaboration = "LIGO Scientific, Virgo",
    title = "{GWTC-2: Compact Binary Coalescences Observed by LIGO and Virgo During the First Half of the Third Observing Run}",
    eprint = "2010.14527",
    archivePrefix = "arXiv",
    primaryClass = "gr-qc",
    reportNumber = "P2000061",
    doi = "10.1103/PhysRevX.11.021053",
    journal = "Phys. Rev. X",
    volume = "11",
    pages = "021053",
    year = "2021"
}

@article{LIGOScientific:2021usb,
    author = "Abbott, R. and others",
    collaboration = "LIGO Scientific, VIRGO",
    title = "{GWTC-2.1: Deep extended catalog of compact binary coalescences observed by LIGO and Virgo during the first half of the third observing run}",
    eprint = "2108.01045",
    archivePrefix = "arXiv",
    primaryClass = "gr-qc",
    reportNumber = "LIGO-P2100063",
    doi = "10.1103/PhysRevD.109.022001",
    journal = "Phys. Rev. D",
    volume = "109",
    number = "2",
    pages = "022001",
    year = "2024"
}

@article{KAGRA:2021vkt,
    author = "Abbott, R. and others",
    collaboration = "KAGRA, VIRGO, LIGO Scientific",
    title = "{GWTC-3: Compact Binary Coalescences Observed by LIGO and Virgo during the Second Part of the Third Observing Run}",
    eprint = "2111.03606",
    archivePrefix = "arXiv",
    primaryClass = "gr-qc",
    reportNumber = "LIGO-P2000318",
    doi = "10.1103/PhysRevX.13.041039",
    journal = "Phys. Rev. X",
    volume = "13",
    number = "4",
    pages = "041039",
    year = "2023"
}

@article{LIGOScientific:2026tep,
    author = "Abac, A. G. and others",
    collaboration = "LIGO Scientific, KAGRA",
    title = "{GWTC-4.0: Updating the Gravitational-wave Transient Catalog with Observations from the First Part of the Fourth LIGO{\textendash}Virgo{\textendash}KAGRA Observing Run}",
    doi = "10.3847/2041-8213/ae2c74",
    journal = "Astrophys. J. Lett.",
    volume = "1004",
    number = "2",
    pages = "L22",
    year = "2026"
}

@article{LIGOScientific:2026wfs,
    author = "Abac, None and others",
    collaboration = "LIGO Scientific, VIRGO, KAGRA",
    title = "{GWTC-5.0: Observations from the Second Part of the Fourth LIGO-Virgo-KAGRA Observing Run and Updates to the Gravitational-Wave Transient Catalog}",
    eprint = "2605.27225",
    archivePrefix = "arXiv",
    primaryClass = "gr-qc",
    reportNumber = "LIGO-P2600152",
    month = "5",
    year = "2026",
    journal = ""
}

@article{LIGOScientific:2020kqk,
    author = "Abbott, R. and others",
    collaboration = "LIGO Scientific, Virgo",
    title = "{Population Properties of Compact Objects from the Second LIGO-Virgo Gravitational-Wave Transient Catalog}",
    eprint = "2010.14533",
    archivePrefix = "arXiv",
    primaryClass = "astro-ph.HE",
    reportNumber = "LIGO-P2000077",
    doi = "10.3847/2041-8213/abe949",
    journal = "Astrophys. J. Lett.",
    volume = "913",
    number = "1",
    pages = "L7",
    year = "2021"
}

@article{KAGRA:2021duu,
    author = "Abbott, R. and others",
    collaboration = "KAGRA, VIRGO, LIGO Scientific",
    title = "{Population of Merging Compact Binaries Inferred Using Gravitational Waves through GWTC-3}",
    eprint = "2111.03634",
    archivePrefix = "arXiv",
    primaryClass = "astro-ph.HE",
    reportNumber = "LIGO-P2100239 ; Data release: https://zenodo.org/record/5655785, LIGO-P2100239",
    doi = "10.1103/PhysRevX.13.011048",
    journal = "Phys. Rev. X",
    volume = "13",
    number = "1",
    pages = "011048",
    year = "2023"
}

@article{LIGOScientific:2025pvj,
    author = "Abac, A. G. and others",
    collaboration = "LIGO Scientific, VIRGO, Virgo,, KAGRA",
    title = "{GWTC-4.0: Population Properties of Merging Compact Binaries}",
    eprint = "2508.18083",
    archivePrefix = "arXiv",
    primaryClass = "astro-ph.HE",
    reportNumber = "LIGO-P2400004",
    doi = "10.3847/2041-8213/ae771e",
    journal = "Astrophys. J. Lett.",
    volume = "1005",
    number = "2",
    pages = "L51",
    year = "2026"
}

@article{LIGOScientific:2026ctl,
    author = "Abac, None and others",
    collaboration = "LIGO Scientific, VIRGO, KAGRA",
    title = "{GWTC-5.0: Population Properties of Merging Compact Binaries}",
    eprint = "2605.27226",
    archivePrefix = "arXiv",
    primaryClass = "astro-ph.HE",
    reportNumber = "LIGO-P2600045",
    month = "5",
    year = "2026",
    journal = ""
}

@article{Biscoveanu:2026ikx,
    author = "Biscoveanu, Sylvia",
    title = "{The first decade of gravitational-wave measurements of black hole spins}",
    eprint = "2606.06209",
    archivePrefix = "arXiv",
    primaryClass = "gr-qc",
    reportNumber = "LIGO-P2600283",
    month = "6",
    year = "2026"
}

@article{Mandel:2018hfr,
    author = "Mandel, Ilya and Farmer, Alison",
    title = "{Merging stellar-mass binary black holes}",
    eprint = "1806.05820",
    archivePrefix = "arXiv",
    primaryClass = "astro-ph.HE",
    doi = "10.1016/j.physrep.2022.01.003",
    journal = "Phys. Rept.",
    volume = "955",
    pages = "1--24",
    year = "2022"
}

@article{Mapelli:2020vfa,
    author = "Mapelli, Michela",
    title = "{Binary Black Hole Mergers: Formation and Populations}",
    eprint = "2105.12455",
    archivePrefix = "arXiv",
    primaryClass = "astro-ph.HE",
    doi = "10.3389/fspas.2020.00038",
    journal = "Front. Astron. Space Sci.",
    volume = "7",
    pages = "38",
    year = "2020"
}

@inbook{Mapelli:2021taw,
    author = "Mapelli, Michela",
    title = "{Formation Channels of Single and Binary Stellar-Mass Black Holes}",
    eprint = "2106.00699",
    archivePrefix = "arXiv",
    primaryClass = "astro-ph.HE",
    doi = "10.1007/978-981-15-4702-7_16-1",
    year = "2021"
}

@article{Callister:2024cdx,
    author = "Callister, T. A.",
    title = "{Observed Gravitational-Wave Populations}",
    eprint = "2410.19145",
    archivePrefix = "arXiv",
    primaryClass = "astro-ph.HE",
    month = "10",
    year = "2024"
}

@article{Talbot:2018cva,
    author = "Talbot, Colm and Thrane, Eric",
    title = "{Measuring the binary black hole mass spectrum with an astrophysically motivated parameterization}",
    eprint = "1801.02699",
    archivePrefix = "arXiv",
    primaryClass = "astro-ph.HE",
    doi = "10.3847/1538-4357/aab34c",
    journal = "Astrophys. J.",
    volume = "856",
    number = "2",
    pages = "173",
    year = "2018"
}

@article{Foreman-Mackey:2012any,
    author = "Foreman-Mackey, Daniel and Hogg, David W. and Lang, Dustin and Goodman, Jonathan",
    title = "{emcee: The MCMC Hammer}",
    eprint = "1202.3665",
    archivePrefix = "arXiv",
    primaryClass = "astro-ph.IM",
    doi = "10.1086/670067",
    journal = "Publ. Astron. Soc. Pac.",
    volume = "125",
    pages = "306--312",
    year = "2013"
}

@article{Thrane:2018qnx,
    author = "Thrane, Eric and Talbot, Colm",
    title = "{An introduction to Bayesian inference in gravitational-wave astronomy: parameter estimation, model selection, and hierarchical models}",
    eprint = "1809.02293",
    archivePrefix = "arXiv",
    primaryClass = "astro-ph.IM",
    doi = "10.1017/pasa.2019.2",
    journal = "Publ. Astron. Soc. Austral.",
    volume = "36",
    pages = "e010",
    year = "2019",
    note = "[Erratum: Publ.Astron.Soc.Austral. 37, e036 (2020)]"
}

@article{Mandel:2018mve,
    author = "Mandel, Ilya and Farr, Will M. and Gair, Jonathan R.",
    title = "{Extracting distribution parameters from multiple uncertain observations with selection biases}",
    eprint = "1809.02063",
    archivePrefix = "arXiv",
    primaryClass = "physics.data-an",
    doi = "10.1093/mnras/stz896",
    journal = "Mon. Not. Roy. Astron. Soc.",
    volume = "486",
    number = "1",
    pages = "1086--1093",
    year = "2019"
}

@article{Vitale:2020aaz,
    author = "Vitale, Salvatore and Gerosa, Davide and Farr, Will M. and Taylor, Stephen R.",
    title = "{Inferring the properties of a population of compact binaries in presence of selection effects}",
    eprint = "2007.05579",
    archivePrefix = "arXiv",
    primaryClass = "astro-ph.IM",
    doi = "10.1007/978-981-15-4702-7\_45-1",
    journal = "",
    month = "7",
    year = "2020"
}

@article{Allen:1997ad,
    author = "Allen, Bruce and Romano, Joseph D.",
    title = "{Detecting a stochastic background of gravitational radiation: Signal processing strategies and sensitivities}",
    eprint = "gr-qc/9710117",
    archivePrefix = "arXiv",
    reportNumber = "WISC-MILW-97-TH-14",
    doi = "10.1103/PhysRevD.59.102001",
    journal = "Phys. Rev. D",
    volume = "59",
    pages = "102001",
    year = "1999"
}

@article{Romano:2016dpx,
    author = "Romano, Joseph D. and Cornish, Neil J.",
    title = "{Detection methods for stochastic gravitational-wave backgrounds: a unified treatment}",
    eprint = "1608.06889",
    archivePrefix = "arXiv",
    primaryClass = "gr-qc",
    doi = "10.1007/s41114-017-0004-1",
    journal = "Living Rev. Rel.",
    volume = "20",
    number = "1",
    pages = "2",
    year = "2017"
}

@article{KAGRA:2021kbb,
    author = "Abbott, R. and others",
    collaboration = "KAGRA, Virgo, LIGO Scientific",
    title = "{Upper limits on the isotropic gravitational-wave background from Advanced LIGO and Advanced Virgo\textquoteright{}s third observing run}",
    eprint = "2101.12130",
    archivePrefix = "arXiv",
    primaryClass = "gr-qc",
    reportNumber = "LIGO-DCC-P2000314",
    doi = "10.1103/PhysRevD.104.022004",
    journal = "Phys. Rev. D",
    volume = "104",
    number = "2",
    pages = "022004",
    year = "2021"
}

@article{LIGOScientific:2025kry,
    author = "Abac, A. G. and others",
    collaboration = "LIGO Scientific, VIRGO, KAGRA",
    title = "{Cosmological and High Energy Physics implications from gravitational-wave background searches in LIGO-Virgo-KAGRA's O1-O4a runs}",
    eprint = "2510.26848",
    archivePrefix = "arXiv",
    primaryClass = "gr-qc",
    reportNumber = "LIGO-PP2500150",
    month = "10",
    year = "2025"
}

@article{LIGOScientific:2025bgj,
    author = "Abac, A. G. and others",
    collaboration = "LIGO Scientific Collaboration, the Virgo Collaboration,, KAGRA, LIGO Scientific",
    title = "{Upper limits on the isotropic gravitational-wave background from the first part of LIGO, Virgo, and KAGRA{\textquoteright}s fourth observing run}",
    eprint = "2508.20721",
    archivePrefix = "arXiv",
    primaryClass = "gr-qc",
    reportNumber = "LIGO-P2500349",
    doi = "10.1103/wq57-sjt2",
    journal = "Phys. Rev. D",
    volume = "114",
    number = "4",
    pages = "042001",
    year = "2026"
}

@article{Zhu:2011bd,
    author = "Zhu, Xing-Jiang and Howell, E. and Regimbau, T. and Blair, D. and Zhu, Zong-Hong",
    title = "{Stochastic Gravitational Wave Background from Coalescing Binary Black Holes}",
    eprint = "1104.3565",
    archivePrefix = "arXiv",
    primaryClass = "gr-qc",
    reportNumber = "LIGO-P1000181-V3",
    doi = "10.1088/0004-637X/739/2/86",
    journal = "Astrophys. J.",
    volume = "739",
    pages = "86",
    year = "2011"
}

@article{Mandic:2012pj,
    author = "Mandic, V. and Thrane, E. and Giampanis, S. and Regimbau, T.",
    title = "{Parameter Estimation in Searches for the Stochastic Gravitational-Wave Background}",
    eprint = "1209.3847",
    archivePrefix = "arXiv",
    primaryClass = "astro-ph.CO",
    doi = "10.1103/PhysRevLett.109.171102",
    journal = "Phys. Rev. Lett.",
    volume = "109",
    pages = "171102",
    year = "2012"
}

@article{Suvodeep:2019,
    author = "Mukherjee, Suvodip and Silk, Joseph",
    title = "{Time-dependence of the astrophysical stochastic gravitational wave background}",
    eprint = "1912.07657",
    archivePrefix = "arXiv",
    primaryClass = "gr-qc",
    doi = "10.1093/mnras/stz3226",
    journal = "Mon. Not. Roy. Astron. Soc.",
    volume = "491",
    number = "4",
    pages = "4690--4701",
    year = "2020"
}

@article{Sah:2025agw,
    author = "Sah, Mohit Raj and Mukherjee, Suvodip",
    title = "{First upper bound for the nonstationary gravitational wave background and its implication for the high redshift binary black hole merger rate}",
    eprint = "2511.03262",
    archivePrefix = "arXiv",
    primaryClass = "astro-ph.HE",
    doi = "10.1103/fzjl-vt37",
    journal = "Phys. Rev. D",
    volume = "113",
    number = "12",
    pages = "123058",
    year = "2026"
}

@article{Bavera:2021wmw,
    author = "Bavera, Simone S. and Franciolini, Gabriele and Cusin, Giulia and Riotto, Antonio and Zevin, Michael and Fragos, Tassos",
    title = "{Stochastic gravitational-wave background as a tool for investigating multi-channel astrophysical and primordial black-hole mergers}",
    eprint = "2109.05836",
    archivePrefix = "arXiv",
    primaryClass = "astro-ph.CO",
    doi = "10.1051/0004-6361/202142208",
    journal = "Astron. Astrophys.",
    volume = "660",
    pages = "A26",
    year = "2022"
}

@article{Sah:2023bgr,
    author = "Sah, Mohit Raj and Mukherjee, Suvodip",
    title = "{Non-stationary astrophysical stochastic gravitational-wave background: a new probe to the high-redshift population of binary black holes}",
    eprint = "2307.06405",
    archivePrefix = "arXiv",
    primaryClass = "gr-qc",
    doi = "10.1093/mnras/stad3365",
    journal = "Mon. Not. Roy. Astron. Soc.",
    volume = "527",
    number = "2",
    pages = "4100--4111",
    year = "2023"
}

@article{Turbang:2023tjk,
    author = "Turbang, Kevin and Lalleman, Max and Callister, Thomas A. and van Remortel, Nick",
    title = "{The Metallicity Dependence and Evolutionary Times of Merging Binary Black Holes: Combined Constraints from Individual Gravitational-wave Detections and the Stochastic Background}",
    eprint = "2310.17625",
    archivePrefix = "arXiv",
    primaryClass = "astro-ph.HE",
    doi = "10.3847/1538-4357/ad3d5c",
    journal = "Astrophys. J.",
    volume = "967",
    number = "2",
    pages = "142",
    year = "2024"
}

@article{Lalleman:2024zjt,
    author = "Lalleman, Max and Turbang, Kevin and Callister, Thomas A. and Van Remortel, Nick",
    title = "{Estimating the redshift dependence of the BBH population using joint CBC and GWB analysis}",
    doi = "10.22323/1.441.0114",
    journal = "PoS",
    volume = "TAUP2023",
    pages = "114",
    year = "2024"
}

@article{Kou:2024gvp,
    author = "Kou, Xiao-Xiao and Fragione, Giacomo and Mandic, Vuk",
    title = "{Stochastic gravitational wave background from binary black hole mergers dynamically assembled in dense star clusters}",
    eprint = "2401.04347",
    archivePrefix = "arXiv",
    primaryClass = "gr-qc",
    doi = "10.1103/PhysRevD.109.123036",
    journal = "Phys. Rev. D",
    volume = "109",
    number = "12",
    pages = "123036",
    year = "2024"
}

@article{Madau:2016jbv,
    author = "Madau, Piero and Fragos, Tassos",
    title = "{Radiation Backgrounds at Cosmic Dawn: X-Rays from Compact Binaries}",
    eprint = "1606.07887",
    archivePrefix = "arXiv",
    primaryClass = "astro-ph.GA",
    doi = "10.3847/1538-4357/aa6af9",
    journal = "Astrophys. J.",
    volume = "840",
    number = "1",
    pages = "39",
    year = "2017"
}

@article{Wu:2011ac,
    author = "Wu, C. and Mandic, V. and Regimbau, T.",
    title = "{Accessibility of the Gravitational-Wave Background due to Binary Coalescences to Second and Third Generation Gravitational-Wave Detectors}",
    eprint = "1112.1898",
    archivePrefix = "arXiv",
    primaryClass = "gr-qc",
    doi = "10.1103/PhysRevD.85.104024",
    journal = "Phys. Rev. D",
    volume = "85",
    pages = "104024",
    year = "2012"
}

@article{Ebersold:2025izh,
    author = "Ebersold, Michael and Regimbau, Tania",
    title = "{Uncertainty in predicting the stochastic gravitational wave background from compact binary coalescences}",
    eprint = "2510.02163",
    archivePrefix = "arXiv",
    primaryClass = "gr-qc",
    doi = "10.1103/p3dn-x5sp",
    journal = "Phys. Rev. D",
    volume = "113",
    number = "4",
    pages = "044026",
    year = "2026"
}

@article{Essick:2022ojx,
    author = "Essick, Reed and Farr, Will",
    title = "{Precision Requirements for Monte Carlo Sums within Hierarchical Bayesian Inference}",
    eprint = "2204.00461",
    archivePrefix = "arXiv",
    primaryClass = "astro-ph.IM",
    month = "4",
    year = "2022"
}

@article{Drasco,
    author = "Steve Drasco andEanna E. Flanagan",
    title = "Detection methods for non-Gaussian gravitational wave stochastic backgrounds",
    journal ="Phys. Rev. D",
    volume = 67,
    pages = "082003",
    year = 2003}

@article{Smith:2017vfk,
    author = "Smith, Rory and Thrane, Eric",
    title = "{Optimal Search for an Astrophysical Gravitational-Wave Background}",
    eprint = "1712.00688",
    archivePrefix = "arXiv",
    primaryClass = "gr-qc",
    reportNumber = "LIGO-DOCUMENT-ID-LIGO-P1700407",
    doi = "10.1103/PhysRevX.8.021019",
    journal = "Phys. Rev. X",
    volume = "8",
    number = "2",
    pages = "021019",
    year = "2018"
}

@article{HernandezVivanco:2019fku,
    author = "Hernandez Vivanco, Francisco and Smith, Rory and Thrane, Eric and Lasky, Paul D.",
    title = "{Accelerated detection of the binary neutron star gravitational-wave background}",
    eprint = "1903.05778",
    archivePrefix = "arXiv",
    primaryClass = "gr-qc",
    doi = "10.1103/PhysRevD.100.043023",
    journal = "Phys. Rev. D",
    volume = "100",
    number = "4",
    pages = "043023",
    year = "2019"
}

@article{Johnson:2024foj,
    author = "Johnson, Aaron D. and Chatziioannou, Katerina and Farr, Will M.",
    title = "{Source confusion from neutron star binaries in ground-based gravitational wave detectors is minimal}",
    eprint = "2402.06836",
    archivePrefix = "arXiv",
    primaryClass = "gr-qc",
    doi = "10.1103/PhysRevD.109.084015",
    journal = "Phys. Rev. D",
    volume = "109",
    number = "8",
    pages = "084015",
    year = "2024"
}

@article{Smith:2020lkj,
    author = "Smith, Rory J. E. and Talbot, Colm and Hernandez Vivanco, Francisco and Thrane, Eric",
    title = "{Inferring the population properties of binary black holes from unresolved gravitational waves}",
    eprint = "2004.09700",
    archivePrefix = "arXiv",
    primaryClass = "astro-ph.HE",
    doi = "10.1093/mnras/staa1642",
    journal = "Mon. Not. Roy. Astron. Soc.",
    volume = "496",
    number = "3",
    pages = "3281--3290",
    year = "2020"
}

@article{Biscoveanu:2020gds,
    author = "Biscoveanu, Sylvia and Talbot, Colm and Thrane, Eric and Smith, Rory",
    title = "{Measuring the primordial gravitational-wave background in the presence of astrophysical foregrounds}",
    eprint = "2009.04418",
    archivePrefix = "arXiv",
    primaryClass = "astro-ph.HE",
    reportNumber = "LIGO Document Number LIGO-P2000297",
    doi = "10.1103/PhysRevLett.125.241101",
    journal = "Phys. Rev. Lett.",
    volume = "125",
    pages = "241101",
    year = "2020"
}

@article{Banagiri:2020kqd,
    author = "Banagiri, Sharan and Mandic, Vuk and Scarlata, Claudia and Yang, Kate Z.",
    title = "{Measuring angular N-point correlations of binary black hole merger gravitational-wave events with hierarchical Bayesian inference}",
    eprint = "2006.00633",
    archivePrefix = "arXiv",
    primaryClass = "astro-ph.CO",
    doi = "10.1103/PhysRevD.102.063007",
    journal = "Phys. Rev. D",
    volume = "102",
    number = "6",
    pages = "063007",
    year = "2020"
}

@article{Talbot:2021igi,
    author = "Talbot, Colm and Thrane, Eric and Biscoveanu, Sylvia and Smith, Rory",
    title = "{Inference with finite time series: Observing the gravitational Universe through windows}",
    eprint = "2106.13785",
    archivePrefix = "arXiv",
    primaryClass = "astro-ph.IM",
    doi = "10.1103/PhysRevResearch.3.043049",
    journal = "Phys. Rev. Res.",
    volume = "3",
    number = "4",
    pages = "043049",
    year = "2021"
}

@article{Talbot:2020auc,
    author = "Talbot, Colm and Thrane, Eric",
    title = "{Gravitational-wave astronomy with an uncertain noise power spectral density}",
    eprint = "2006.05292",
    archivePrefix = "arXiv",
    primaryClass = "astro-ph.IM",
    doi = "10.1103/PhysRevResearch.2.043298",
    journal = "Phys. Rev. Res.",
    volume = "2",
    number = "4",
    pages = "043298",
    year = "2020"
}

@article{Kou:2025bhk,
    author = "Kou, Xiao-Xiao and Saleem, Muhammed and Mandic, Vuk and Talbot, Colm and Thrane, Eric",
    title = "{Progress toward the detection of the gravitational-wave background from stellar-mass binary black holes: A mock data challenge}",
    eprint = "2506.14179",
    archivePrefix = "arXiv",
    primaryClass = "gr-qc",
    doi = "10.1103/h9w1-94m1",
    journal = "Phys. Rev. D",
    volume = "112",
    number = "8",
    pages = "084064",
    year = "2025"
}

@article{Talbot:2025vth,
    author = "Talbot, Colm and others",
    title = "{Inference with finite time series: II. The window strikes back}",
    eprint = "2508.11091",
    archivePrefix = "arXiv",
    primaryClass = "gr-qc",
    doi = "10.1088/1361-6382/ae1ac7",
    journal = "Class. Quant. Grav.",
    volume = "42",
    number = "23",
    pages = "235023",
    year = "2025"
}

@article{Bers:2025tei,
    author = "Bers, Nico and Biscoveanu, Sylvia",
    title = "{Probing the Peak of Star Formation with the Stochastic Background of Binary Black Hole Mergers}",
    eprint = "2506.21868",
    archivePrefix = "arXiv",
    primaryClass = "astro-ph.HE",
    reportNumber = "LIGO document number LIGO-P2500381",
    doi = "10.3847/1538-4357/ae2319",
    journal = "Astrophys. J.",
    volume = "997",
    number = "1",
    pages = "108",
    year = "2026"
}

@article{Finn:1992xs,
    author = "Finn, Lee Samuel and Chernoff, David F.",
    title = "{Observing binary inspiral in gravitational radiation: One interferometer}",
    eprint = "gr-qc/9301003",
    archivePrefix = "arXiv",
    reportNumber = "PRINT-93-0138 (NORTHWESTERN)",
    doi = "10.1103/PhysRevD.47.2198",
    journal = "Phys. Rev. D",
    volume = "47",
    pages = "2198--2219",
    year = "1993"
}

@article{Renzini:2024hiu,
    author = "Renzini, Arianna I. and Callister, Tom and Chatziioannou, Katerina and Farr, Will M.",
    title = "{Background information: A study on the sensitivity of astrophysical gravitational-wave background searches}",
    eprint = "2403.14793",
    archivePrefix = "arXiv",
    primaryClass = "astro-ph.HE",
    doi = "10.1103/PhysRevD.110.023014",
    journal = "Phys. Rev. D",
    volume = "110",
    number = "2",
    pages = "023014",
    year = "2024"
}

@article{Lawrence:2023buo,
    author = "Lawrence, Jessica and Turbang, Kevin and Matas, Andrew and Renzini, Arianna I. and van Remortel, Nick and Romano, Joseph D.",
    title = "{A stochastic search for intermittent gravitational-wave backgrounds}",
    eprint = "2301.07675",
    archivePrefix = "arXiv",
    primaryClass = "gr-qc",
    doi = "10.1103/PhysRevD.107.103026",
    journal = "Phys. Rev. D",
    volume = "107",
    number = "10",
    pages = "103026",
    year = "2023"
}

@article{Liu:2026xhc,
    author = "Liu, Xiaolin and Kuroyanagi, Sachiko",
    title = "{Analyzing intermittent stochastic gravitational wave background I:Effect of detector response}",
    eprint = "2601.10428",
    archivePrefix = "arXiv",
    primaryClass = "gr-qc",
    reportNumber = "IFT-UAM/CSIC-26-6",
    month = "1",
    year = "2026"
}

@article{Kullback:1951zyt,
    author = "Kullback, S. and Leibler, R. A.",
    title = "{On Information and Sufficiency}",
    doi = "10.1214/aoms/1177729694",
    journal = "The Annals of Mathematical Statistics",
    volume = "22",
    number = "1",
    pages = "79--86",
    year = "1951"
}

@article{Skilling:2006gxv,
    author = "Skilling, John",
    title = "{Nested sampling for general Bayesian computation}",
    doi = "10.1214/06-BA127",
    journal = "Bayesian Analysis",
    volume = "1",
    number = "4",
    pages = "833--859",
    year = "2006"
}

@article{Skilling:2004pqw,
    author = "Skilling, John",
    title = "{Nested Sampling}",
    doi = "10.1063/1.1835238",
    journal = "AIP Conf. Proc.",
    volume = "735",
    number = "1",
    pages = "395",
    year = "2004"
}

@article{Higson2017SamplingEI,
  title={Sampling Errors in Nested Sampling Parameter Estimation},
  author={Edward Higson and Will Handley and Michael P. Hobson and Anthony N. Lasenby},
  journal={Bayesian Analysis},
  year={2017},
  eprint = "1703.09701",
  archivePrefix = "arxiv",
  primaryClass = "stat.ME",
  doi = "10.1214/17-BA1075"
}

@article{Higson:2018cqj,
    author = "Higson, Edward and Handley, Will and Hobson, Mike and Lasenby, Anthony",
    title = "{Nestcheck: diagnostic tests for nested sampling calculations}",
    eprint = "1804.06406",
    archivePrefix = "arXiv",
    primaryClass = "stat.CO",
    doi = "10.1093/mnras/sty3090",
    journal = "Mon. Not. Roy. Astron. Soc.",
    volume = "483",
    number = "2",
    pages = "2044--2056",
    year = "2019"
}

@article{Buchner:2014nha,
    author = "Buchner, J. and Georgakakis, A. and Nandra, K. and Hsu, L. and Rangel, C. and Brightman, M. and Merloni, A. and Salvato, M. and Donley, J. and Kocevski, D.",
    title = "{X-ray spectral modelling of the AGN obscuring region in the CDFS: Bayesian model selection and catalogue}",
    eprint = "1402.0004",
    archivePrefix = "arXiv",
    primaryClass = "astro-ph.HE",
    doi = "10.1051/0004-6361/201322971",
    journal = "Astron. Astrophys.",
    volume = "564",
    pages = "A125",
    year = "2014"
}

@article{Handley:2015vkr,
    author = "Handley, W. J. and Hobson, M. P. and Lasenby, A. N.",
    title = "{polychord: next-generation nested sampling}",
    eprint = "1506.00171",
    archivePrefix = "arXiv",
    primaryClass = "astro-ph.IM",
    doi = "10.1093/mnras/stv1911",
    journal = "Mon. Not. Roy. Astron. Soc.",
    volume = "453",
    number = "4",
    pages = "4385--4399",
    year = "2015"
}

@article{Speagle:2019ivv,
    author = "Speagle, Joshua S.",
    title = "{dynesty: a dynamic nested sampling package for estimating Bayesian posteriors and evidences}",
    eprint = "1904.02180",
    archivePrefix = "arXiv",
    primaryClass = "astro-ph.IM",
    doi = "10.1093/mnras/staa278",
    journal = "Mon. Not. Roy. Astron. Soc.",
    volume = "493",
    number = "3",
    pages = "3132--3158",
    year = "2020"
}

@article{Williams:2021qyt,
    author = "Williams, Michael J. and Veitch, John and Messenger, Chris",
    title = "{Nested sampling with normalizing flows for gravitational-wave inference}",
    eprint = "2102.11056",
    archivePrefix = "arXiv",
    primaryClass = "gr-qc",
    doi = "10.1103/PhysRevD.103.103006",
    journal = "Phys. Rev. D",
    volume = "103",
    number = "10",
    pages = "103006",
    year = "2021"
}

@article{Ashton:2022grj,
    author = "Ashton, Greg and others",
    title = "{Nested sampling for physical scientists}",
    eprint = "2205.15570",
    archivePrefix = "arXiv",
    primaryClass = "stat.CO",
    doi = "10.1038/s43586-022-00121-x",
    journal = "Nature",
    volume = "2",
    year = "2022"
}

@article{Fowlie:2022jls,
    author = "Fowlie, Andrew and Li, Qiao and Lv, Huifang and Sun, Yecheng and Zhang, Jia and Zheng, Le",
    title = "{Nested sampling statistical errors}",
    eprint = "2211.03258",
    archivePrefix = "arXiv",
    primaryClass = "astro-ph.IM",
    doi = "10.1093/mnras/stad751",
    journal = "Mon. Not. Roy. Astron. Soc.",
    volume = "521",
    number = "3",
    pages = "4100--4108",
    year = "2023"
}

@article{Karamanis:2022ksp,
    author = "Karamanis, Minas and Nabergoj, David and Beutler, Florian and Peacock, John A. and Seljak, Uros",
    title = "{pocoMC: A Python package for accelerated Bayesian inference in astronomy and cosmology}",
    eprint = "2207.05660",
    archivePrefix = "arXiv",
    primaryClass = "astro-ph.IM",
    doi = "10.21105/joss.04634",
    journal = "J. Open Source Softw.",
    volume = "7",
    number = "79",
    pages = "4634",
    year = "2022"
}

@article{Pratten:2020fqn,
    author = "Pratten, Geraint and Husa, Sascha and Garcia-Quiros, Cecilio and Colleoni, Marta and Ramos-Buades, Antoni and Estelles, Hector and Jaume, Rafel",
    title = "{Setting the cornerstone for a family of models for gravitational waves from compact binaries: The dominant harmonic for nonprecessing quasicircular black holes}",
    eprint = "2001.11412",
    archivePrefix = "arXiv",
    primaryClass = "gr-qc",
    reportNumber = "LIGO-P2000018",
    doi = "10.1103/PhysRevD.102.064001",
    journal = "Phys. Rev. D",
    volume = "102",
    number = "6",
    pages = "064001",
    year = "2020"
}

@article{Iacovelli:2022bbs,
    author = "Iacovelli, Francesco and Mancarella, Michele and Foffa, Stefano and Maggiore, Michele",
    title = "{Forecasting the Detection Capabilities of Third-generation Gravitational-wave Detectors Using GWFAST}",
    eprint = "2207.02771",
    archivePrefix = "arXiv",
    primaryClass = "gr-qc",
    doi = "10.3847/1538-4357/ac9cd4",
    journal = "Astrophys. J.",
    volume = "941",
    number = "2",
    pages = "208",
    year = "2022"
}

@misc{LIGO-T2200043,
  note = {\url{https://dcc.ligo.org/T2200043-v3/public}}
}

@article{Sasli:2026pds,
    author = "Sasli, Argyro and Karamanis, Minas and Karnesis, Nikolaos and Coughlin, Michael W. and Mandic, Vuk and Seljak, Uro{\v{s}} and Stergioulas, Nikolaos",
    title = "{Beyond Gaussian assumptions: A new robust statistical framework for gravitational-wave data analysis}",
    eprint = "2602.22074",
    archivePrefix = "arXiv",
    primaryClass = "gr-qc",
    doi = "10.1103/n5vd-kzp1",
    journal = "Phys. Rev. D",
    volume = "114",
    number = "2",
    pages = "024004",
    year = "2026"
}

@article{Sasli:2023mxr,
    author = "Sasli, Argyro and Karnesis, Nikolaos and Stergioulas, Nikolaos",
    title = "{Heavy-tailed likelihoods for robustness against data outliers: Applications to the analysis of gravitational wave data}",
    eprint = "2305.04709",
    archivePrefix = "arXiv",
    primaryClass = "gr-qc",
    doi = "10.1103/PhysRevD.108.103005",
    journal = "Phys. Rev. D",
    volume = "108",
    number = "10",
    pages = "103005",
    year = "2023"
}

@article{Karnesis:2024pxh,
    author = "Karnesis, Nikolaos and Sasli, Argyro and Buscicchio, Riccardo and Stergioulas, Nikolaos",
    title = "{Characterization of non-Gaussian stochastic signals with heavier-tailed likelihoods}",
    eprint = "2410.14354",
    archivePrefix = "arXiv",
    primaryClass = "gr-qc",
    doi = "10.1103/PhysRevD.111.022005",
    journal = "Phys. Rev. D",
    volume = "111",
    number = "2",
    pages = "022005",
    year = "2025"
}

@article{Fishbach:2019ckx,
    author = "Fishbach, Maya and Farr, Will M. and Holz, Daniel E.",
    title = "{The Most Massive Binary Black Hole Detections and the Identification of Population Outliers}",
    eprint = "1911.05882",
    archivePrefix = "arXiv",
    primaryClass = "astro-ph.HE",
    doi = "10.3847/2041-8213/ab77c9",
    journal = "Astrophys. J. Lett.",
    volume = "891",
    number = "2",
    pages = "L31",
    year = "2020"
}

@article{Farah:2023vsc,
    author = "Farah, Amanda M. and Edelman, Bruce and Zevin, Michael and Fishbach, Maya and Ezquiaga, Jose Mar{\'\i}a and Farr, Ben and Holz, Daniel E.",
    title = "{Things That Might Go Bump in the Night: Assessing Structure in the Binary Black Hole Mass Spectrum}",
    eprint = "2301.00834",
    archivePrefix = "arXiv",
    primaryClass = "astro-ph.HE",
    doi = "10.3847/1538-4357/aced02",
    journal = "Astrophys. J.",
    volume = "955",
    number = "2",
    pages = "107",
    year = "2023"
}
\end{document}